\documentclass[useAMS,usenatbib]{mn2e}

\usepackage{verbatim}
\usepackage{graphicx}
\usepackage{epstopdf}
\usepackage{subfigure}
\usepackage{color}

\newcommand{\mdot}{M$_\odot$ yr$^{-1}$}

\newcommand{\apjl}{{ApJ Lett.}}		
\newcommand{\apjs}{{ApJS}}		
		
\newcommand{\aap}{{A\&A.}}

\title{Classical T Tauri stars: magnetic fields, coronae, and star-disc interactions}

\author[Johnstone et al.]{C. P. Johnstone$^{1}$, M. Jardine$^2$, S. G. Gregory$^{2}$, J.-F. Donati$^3$, G. Hussain$^4$\\ $^1$ University of Vienna, Department of Astronomy, T\"{u}rkenschanzstrasse 17, 1180 Vienna, Austria\\ $^2$ School of Physics and Astronomy, Universtiy of St Andrews, St Andrews, Scotland KY16 9SS \\  $^3$ LATT-UMR 5572, CNRS \& Univ. P. Sabatier, 14 Av. E. Belin, Toulouse, F-31400, France\\$^4$ ESO, Karl-Schwarzchild-Str. 2, D-85748 Garching, Germany}

\begin{document}

\maketitle

\begin{abstract}
The magnetic fields of young stars set their coronal properties and control their spin evolution via the star-disc interaction and outflows. 
Using 14 magnetic maps of 10 classical T Tauri stars (CTTSs) we investigate their closed X-ray emitting coronae, their open wind-bearing magnetic fields, and the geometry of magnetospheric accretion flows.  
The magnetic fields of all the CTTSs are multipolar. 
Stars with simpler (more dipolar) large-scale magnetic fields have stronger fields, are slower rotators, and have larger X-ray emitting coronae compared to stars with more complex large-scale magnetic fields.   
The field complexity controls the distribution of open and closed field regions across the stellar surface, and strongly influences the location and shapes of accretion hot spots.  However, the higher order field components are of secondary importance in determining the total unsigned open magnetic flux, which depends mainly on the strength of the dipole component and the stellar
surface area.  
Likewise, the dipole component alone provides an adequate approximation of the disc truncation radius.  
For some stars, the pressure of the hot coronal plasma dominates the stellar magnetic pressure and forces open the closed field inside the disc truncation radius. 
This is significant as accretion models generally assume that the magnetic field has a closed geometry out to the inner disc edge.
\end{abstract}


\section{Introduction}

% Standard intro remarks

T Tauri stars are low-mass, pre-main sequence stars characterised by large irregular variability and the presence of strong emission lines (\mbox{\citealt{1994AJ....108.1056E}}; \mbox{\citealt{2013ApJS..207....1A}}).
They are classified as either classical \mbox{T Tauri} stars (CTTSs) or weak-line \mbox{T Tauri} stars (WTTSs).
CTTSs show excess IR emission from circumstellar discs, and excess UV and optical emission from the accretion of disc material. 
WTTSs are in a more evolved state where they have lost their discs and are no longer accreting.

% Magnetic fields on CTTS

CTTSs host strong magnetic fields with measured surface averaged strengths of up to a few kG (\mbox{\citealt{2007ApJ...664..975J}}).
These fields are responsible for several processes on CTTSs, such as X-ray emission from magnetically confined coronae, the truncation of circumstellar discs several stellar radii from the star, and magnetospheric accretion. 
In order to study these processes, surface averaged magnetic field strengths are not sufficient and it is necessary to know about the topologies of the large-scale magnetic fields of these stars.  
This is done using the Zeeman-Doppler Imaging (ZDI) technique, which uses time-resolved spectropolarimetric observations of stars over several rotations to reconstruct the distributions and orientations of the large-scale magnetic fields over their surfaces (\mbox{\citealt{1989A&A...225..456S}}; \mbox{\citealt{2001LNP...573..207D}}). 

ZDI magnetic maps have recently been published for a sample of CTTSs with different masses and ages as part of the Magnetic Protostars and Planets project (MaPP; \mbox{\citealt{2010MNRAS.409.1347D}}) and have shown that CTTSs possess magnetic fields with a range of strengths and topologies.
The topologies range from simple, almost axisymmetric fields on \mbox{AA Tau} and \mbox{BP Tau}, to much more complex non-axisymmetric fields on \mbox{CR Cha}, \mbox{CV Cha}, and \mbox{V2247 Oph}. 
The strengths and topologies of these fields appear to be related to the internal structures of the host stars (\mbox{\citealt{2012ApJ...755...97G}}).
Stars that are yet to develop radiative cores (e.g. \mbox{AA Tau} and \mbox{BP Tau}) possess strong simple axisymmetric fields dominated by low-order spherical harmonic components.
Stars that are no longer fully convective possess much weaker and more complex non-axisymmetric fields.
The exception is \mbox{V2247 Oph}, which possesses a highly complex and weak field but has too low a mass to develop a radiative core as it evolves.
These results have been reviewed by \mbox{\citet{2011AN....332.1027G}} and \mbox{\citet{2012AN....333....4H}}.
Similar results have been found for the sample of main-sequence M dwarfs that have been studied using ZDI (\mbox{\citealt{2008MNRAS.390..545D}}; \mbox{\citealt{2008MNRAS.390..567M}}; \mbox{\citealt{2010MNRAS.407.2269M}}), though for PMS stars, the situation is more complicated because the internal structure of PMS stars is determined by both mass and age (\mbox{\citealt{2012ApJ...755...97G}}).

% X-ray emission blah blah blah

\mbox{T Tauri} stars show very high levels of X-ray emission, with typical X-ray luminosities of $10^{28}-10^{32}$ erg s$^{-1}$ (\mbox{\citealt{1981ApJ...250..254W}}; \mbox{\citealt{2005ApJS..160..401P}}). 
In most cases, the majority of X-ray emission comes from hot magnetically confined coronae (\mbox{\citealt{2006ApJ...649..914S}}).
Whereas the solar corona is at temperatures of a few MK, the coronae of \mbox{T Tauri} stars can up to 50 MK (\mbox{\citealt{2005ApJS..160..401P}}; \mbox{\citealt{2007A&A...468..353G}}). 
X-ray emission from CTTSs also comes from large stellar flares (e.g. \mbox{\citealt{2008ApJ...688..418G}}), X-ray jets (e.g. \mbox{\citealt{2008A&A...478..797G}}), and accretion shocks where material accreting from the disc impacts the stellar surface at supersonic speeds (e.g. \mbox{\citealt{2011A&A...530A...1A}}).
The most extreme example of X-ray emission from an accretion shock is the case of \mbox{TW Hya}.
\mbox{TW Hya} is viewed almost pole-on, which makes it ideal for observing emission from accretion shock regions.
Analysis of \mbox{TW Hya}'s X-ray spectrum has indicated that the star's X-ray emission measure is dominated by a low-temperature, high density plasma, most likely formed in accretion shocks (\mbox{\citealt{2002ApJ...567..434K}}; \mbox{\citealt{2006A&A...449..737R}}; \mbox{\citealt{2010ApJ...710.1835B}}).

% Accretion stuff

Early disc and accretion models assumed that discs extend all the way to the stellar surface and accretion occurs through a boundary layer (e.g. \mbox{\citealt{1974MNRAS.168..603L}}).
However, the strong magnetic fields of CTTSs are able to truncate the disc far from the stellar surface and funnel accretion along magnetic field lines at approximately free-fall velocities - typically a few hundred \mbox{km s$^{-1}$} - onto the stellar surface.
This model was originally developed for magnetic neutron stars (\mbox{\citealt{1977ApJ...217..578G}}; \mbox{\citealt{1979ApJ...232..259G}}; \mbox{\citealt{1979ApJ...234..296G}}) and applied to CTTSs by \mbox{\citet{1990RvMA....3..234C}} and \mbox{\citet{1991ApJ...370L..39K}}.
When accreting material impacts the stellar surface, accretion shocks are formed.
The post-shock plasma has densities of $10^{12}-10^{13}$ cm$^{-3}$ and temperatures of a few MK (\mbox{\citealt{2011A&A...530A...1A}}).
These shocks account for the excess UV and optical emission from CTTSs (\mbox{\citealt{1991ApJ...370L..39K}}), and are also responsible for significant amounts of softer X-ray emission.

% Modeling coronae from ZDI magnetograms

By making assumptions about the magnetic fields, 2D surface magnetic maps can be used to reconstruct the 3D coronal field geometries.  
These extrapolations can be used to estimate the structure of the X-ray emitting closed corona.
\mbox{\citet{2002MNRAS.336.1364J}} used potential field extrapolations of the young rapidly rotating main-sequence star AB Dor to infer the global X-ray emission measure and rotational modulation.
They found that a magnetically confined isothermal corona in hydrostatic equilibrium could give the observed emission measures with average coronal electron densities of $10^{9.0}-10^{10.7}$ cm$^{-3}$. 
\mbox{\citet{2006MNRAS.367..917J}} used a similar model to show how the differences between the coronal properties of stars are related to their magnetic field geometries. 
They found that stars with complex magnetic fields tend to have coronal plasma densities that are an order of magnitude higher than stars with simple dipolar geometries. 
They also found that the observed correlation between \mbox{X-ray} emission and stellar mass (\mbox{\citealt{2005ApJS..160..401P}}; \mbox{\citealt{2007A&A...468..353G}}) can arise from variations in the surface gravities and the sizes of the closed coronae.

These field extrapolations are also useful for studying magnetospheric accretion onto CTTSs. 
Most accretion models assume that the stellar magnetic field is an axisymmetric dipole.
However, recent studies using ZDI have shown that CTTSs often have much more complex field structures.
For an axisymmetric dipole field, accretion footpoints can be expected to be seen distributed in bands around the star.
However, if the dipole component is tilted with respect to the stars rotation axis, these bands no longer extend all the way around the star, and cover a smaller fraction of the stellar surface (\mbox{\citealt{1998ApJ...497..342M}}; {\citealt{2003ApJ...595.1009R}}; \mbox{\citealt{2004ApJ...610..920R}}; \mbox{\citealt{2006MNRAS.371..999G}}).
\mbox{\citet{2006MNRAS.371..999G}} used field extrapolations of surface fields for real stars to model accretion onto stars with complex magnetic fields. 
For such fields, accretion hotspots can be distributed over the stellar surface in highly complex patterns.
Similar results were found using ZDI magnetic maps from \mbox{BP Tau} and \mbox{V2129 Oph} by \mbox{\citet{2008MNRAS.389.1839G}} and \mbox{\citet{2008MNRAS.386..688J}}, and using 3D MHD simulations by \mbox{\citet{2011MNRAS.413.1061L}} and \mbox{\citet{2011MNRAS.411..915R}}.
They found that the latitudes at which accreting material impacts the stellar surface is determined by the distance from the star that the disc is truncated, with large disc truncation radii leading to accretion located at high latitudes, and small disc truncation radii leading to accretion footpoints distributed over a range of latitudes.

% Summary of paper to come

In this paper, we analyse the magnetic fields of the sample of CTTSs with published ZDI magnetic maps and use field extrapolations to study the closed X-ray emitting coronae, and accretion geometries on these stars.
This is similar to \mbox{\citet{2012MNRAS.424.1077L}} who used similar models to investigate the field structures and coronal X-ray emission on the sample of M dwarf main-sequence stars with published ZDI maps.
In Section \ref{sect:sample}, we discuss the stars in the sample; in Section \ref{sect:surfacemaps}, we analyse the ZDI magnetic maps for these stars; in Section \ref{sect:corona}, we model their closed X-ray emitting coronae; in Section \ref{sect:acc}, we calculate disc truncation radii, and accretion geometries, for these stars; in Section \ref{sect:summary}, we summarise the results, and discuss our conclusions.

\begin{figure*}
\subfigure[AA Tau 2009]{\includegraphics[trim= 0cm 0cm 1cm 0cm, width=0.41\textwidth]{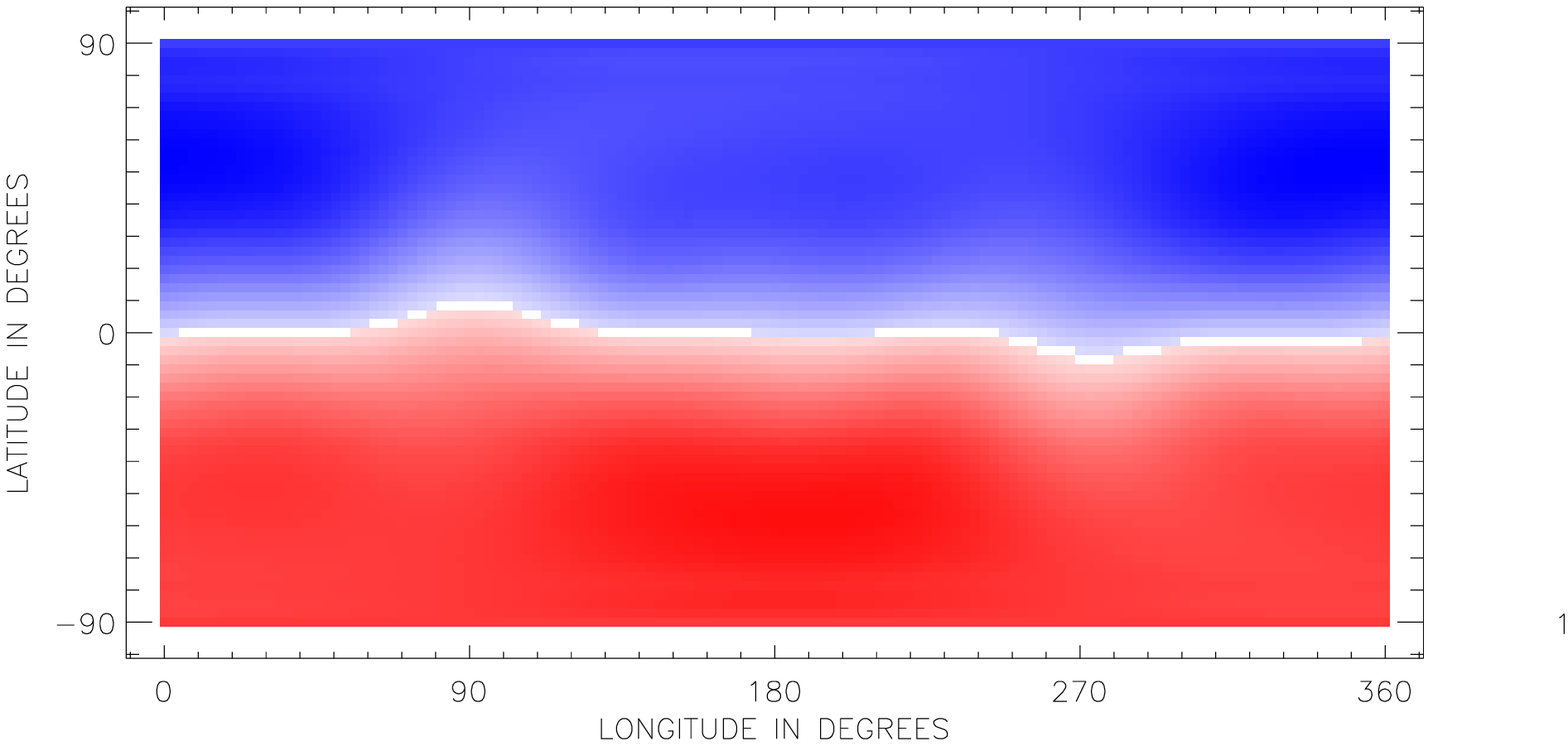}}
\includegraphics[trim=2cm -1.5cm 1.2cm 1cm, clip=true, width=0.08\textwidth]{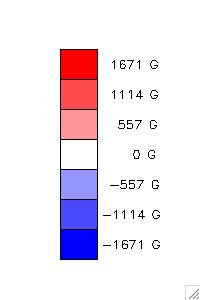}
\subfigure[V2247 Oph]{\includegraphics[trim= 0cm 0cm 1cm 0cm, width=0.41\textwidth]{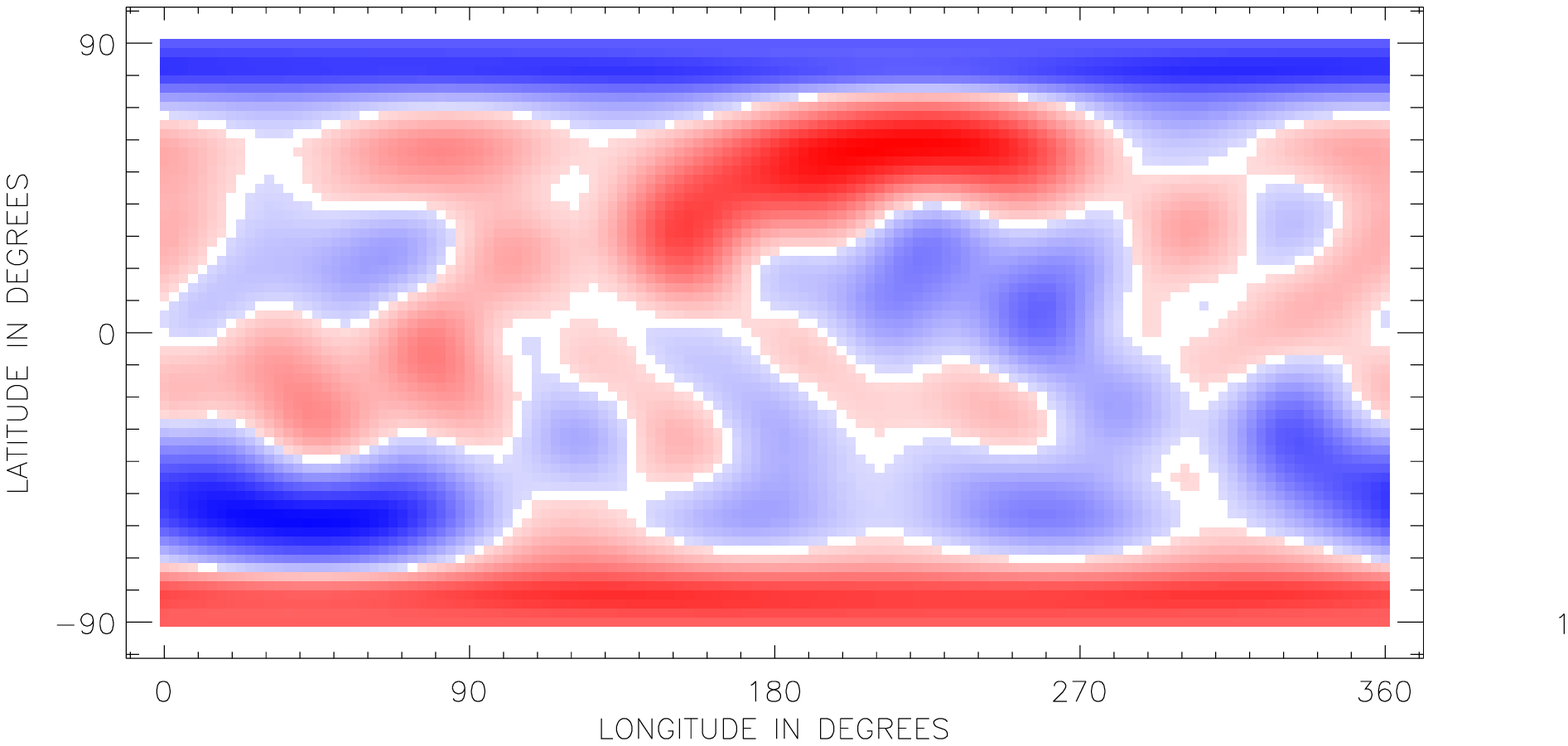}}
\includegraphics[trim=2cm -1.5cm 1.2cm 1cm, clip=true, width=0.08\textwidth]{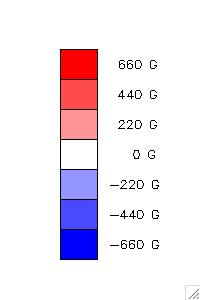}\\
\subfigure[CR Cha]{\includegraphics[trim= 0cm 0cm 1cm 0cm, width=0.41\textwidth]{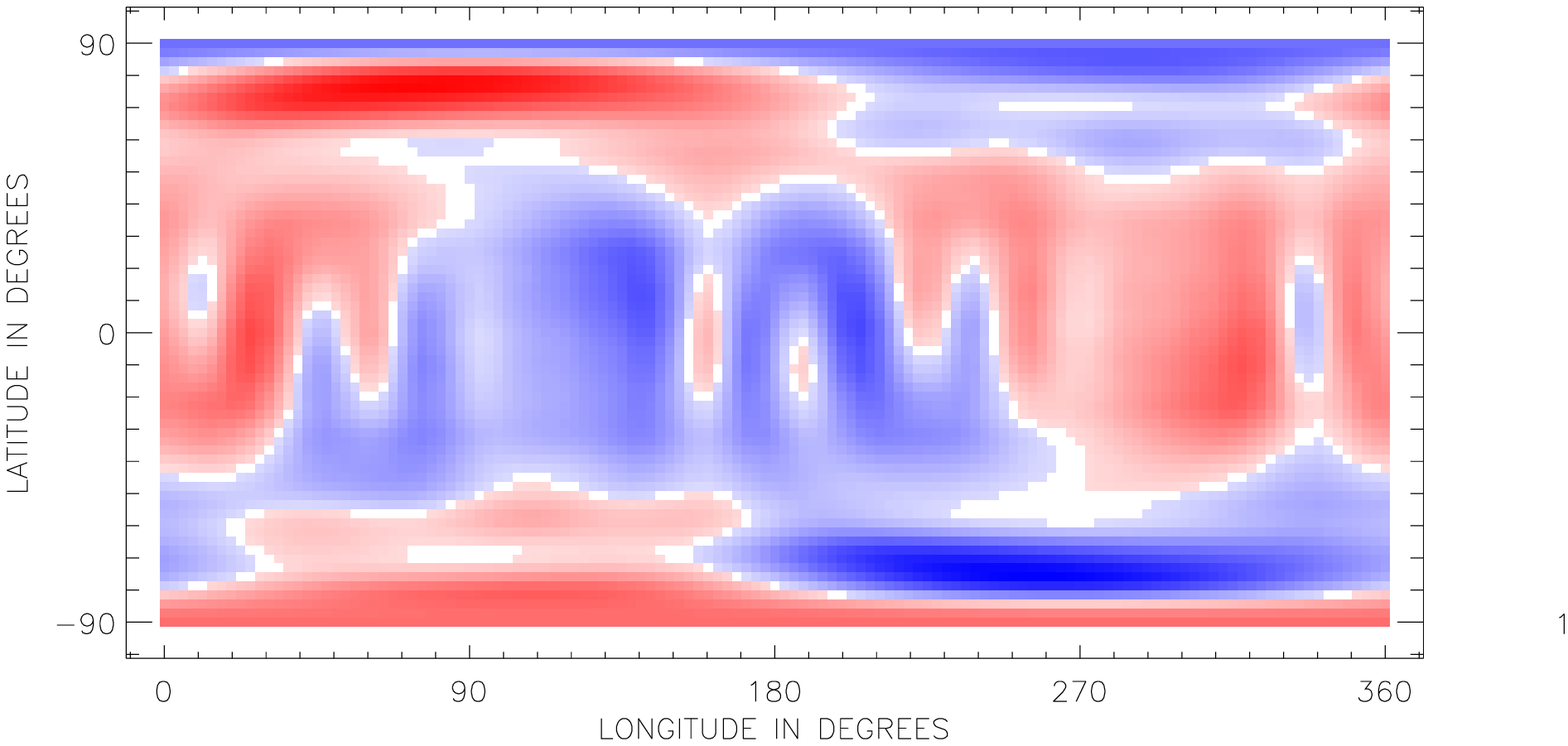}}
\includegraphics[trim=2cm -1.5cm 1.2cm 1cm, clip=true, width=0.08\textwidth]{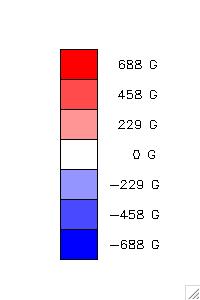}
\subfigure[CV Cha]{\includegraphics[trim= 0cm 0cm 1cm 0cm, width=0.41\textwidth]{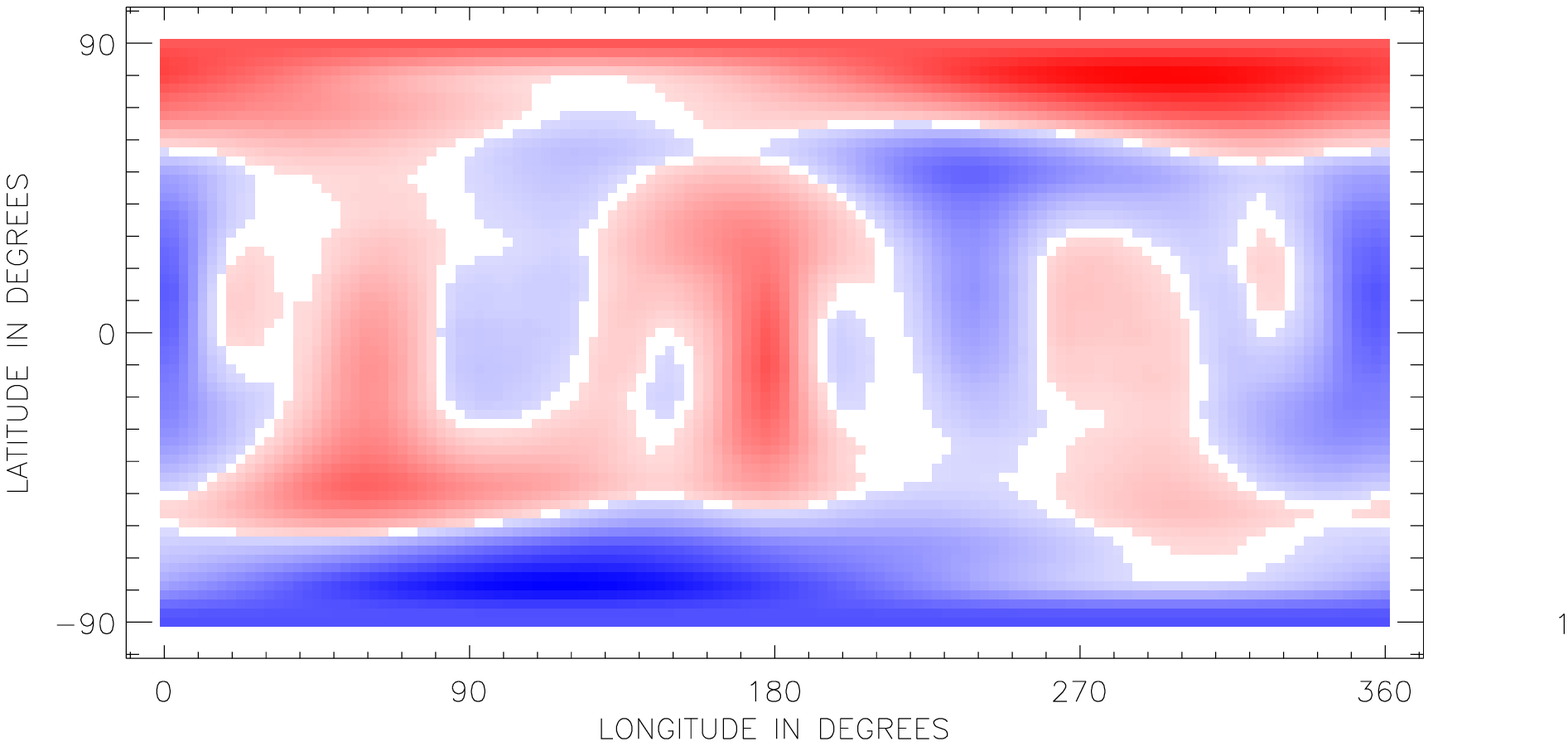}}
\includegraphics[trim=2cm -1.5cm 1.2cm 1cm, clip=true, width=0.08\textwidth]{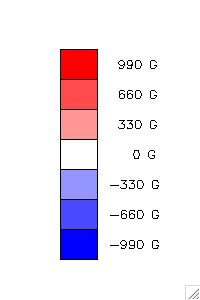}
\caption{
The radial components of the ZDI magnetic maps for \mbox{AA Tau} (\mbox{\citealt{2010MNRAS.409.1347D}}), \mbox{V2247 Oph} (\mbox{\citealt{2010MNRAS.402.1426D}}), \mbox{CR Cha}, and \mbox{CV Cha}. 
We show the \mbox{AA Tau} and \mbox{V2247 Oph} maps because they are the most simple and the most complex large-scale magnetic fields in the sample; we show the \mbox{CR Cha} and \mbox{CV Cha} maps because they have not  previously been published. 
}
 \label{fig:CTTSdispAll}
\end{figure*}

\section{The Sample} \label{sect:sample}

In this paper, we use the published magnetic maps for the CTTSs  \mbox{AA Tau} ({\citealt{2010MNRAS.409.1347D}}), \mbox{BP Tau} ({\citealt{2008MNRAS.386.1234D}}), \mbox{CR Cha} ({\citealt{2009MNRAS.398..189H}}), \mbox{CV Cha} ({\citealt{2009MNRAS.398..189H}}), \mbox{GQ Lup} ({\citealt{2012MNRAS.425.2948D}}), \mbox{TW Hya} ({\citealt{2011MNRAS.417..472D}}), \mbox{V2129 Oph} ({\citealt{2007MNRAS.380.1297D}}; {\citealt{2011MNRAS.412.2454D}}), \mbox{V2247 Oph} ({\citealt{2010MNRAS.402.1426D}}), and both stars in the binary system \mbox{V4046 Sgr} (\mbox{\citealt{2011MNRAS.417.1747D}}).

The original magnetic maps for \mbox{CR Cha} and \mbox{CV Cha}, published by \mbox{\citet{2009MNRAS.398..189H}}, were produced using a different fitting technique than was used to produce the other maps; therefore, we do not use these maps.
Instead, we use new maps produced using the same observational data but with an updated ZDI code that fits the maps directly to the circularly polarised line profiles as a series of spherical harmonic components, which is the technique used to fit the maps for the other stars.
These maps can be seen in \mbox{Fig. \ref{fig:CTTSdispAll}}.
Another difference between the fitting method used for the \mbox{CR Cha} and \mbox{CV Cha} maps and the fitting method used for the other stars is the fact that Zeeman signatures from accretion proxies were used to fit the magnetic maps of the other stars, but were not used for \mbox{CR Cha} and \mbox{CV Cha} because no Zeeman signatures were detected in accretion related emission lines for these stars.
This could potentially remove magnetic energy from the axisymmetric spherical harmonic components of the field, and move magnetic energy to higher $l$-value modes.
The effects of not including the accretion related emission lines in ZDI fits were explored by \mbox{\citet{2007MNRAS.380.1297D}} for the case of \mbox{V2129 Oph}. 

The stellar parameters for these stars are summarised in \mbox{Table \ref{tbl:stellarparams}} and the published magnetic maps for \mbox{AA Tau} and \mbox{V2247 Oph} are also shown in \mbox{Fig. \ref{fig:CTTSdispAll}}.
Detailed discussions of the adopted stellar parameters for each star can be found in the papers (listed above) where the magnetic maps have been published.
As mass accretion can be highly variable in time, in all cases we use mass accretion rates derived from the same observations as the ZDI maps.
The lowest mass star in the sample is \mbox{V2247 Oph}, which has a mass of \mbox{0.36 M$_\odot$} and is therefore unlikely to develop a radiative core at any age. 
Other than \mbox{AA Tau} and \mbox{BP Tau}, which are both still fully convective, the other stars in the sample all host radiative cores. 

For the coronal models in Section \ref{sect:corona}, we require coronal temperatures and X-ray emission measures for each of the stars.
We take from the literature multi-temperature component fits to observed X-ray spectra (for references, see \mbox{Table \ref{tbl:stellarparams}}).
For most of these stars, these fits have low-temperature components that we assume are a result of accretion shocks; we therefore disregard these components.
As our model assumes an isothermal corona, we combine the remaining components to get the coronal temperature and emission measure using

\begin{equation}
EM = \sum_i EM_i
\end{equation}

\begin{equation}
\log T = \frac{\sum_i EM_i \log T_i}{\sum_i EM_i}.
\end{equation}

\noindent For \mbox{AA Tau}, the components in the fit to the X-ray spectrum by \mbox{\citet{2007A&A...468..353G}} are all above 10 MK, and are therefore unlikely to have been produced in accretion shocks; we therefore assume that all of these components correspond entirely to coronal emission. 
For \mbox{BP Tau} and \mbox{CR Cha}, the three temperature component fits given by \mbox{\citet{2006A&A...449..737R}} both have low-temperature components at approximately 2 MK, which could be a result of X-ray emission from accretion shocks; we therefore disregard these components.
However, these components are only 18\% of the total emission measures, so this only makes a small difference to the coronal models. 
Similarly, for \mbox{V2129 Oph}, we disregard the three lowest temperature components of the nine-temperature component emission measure distribution (EMD) fit of \mbox{\citet{2011A&A...530A...1A}}; these components only correspond to 8\% of the total emission measure. 
For \mbox{V2247 Oph}, the low-temperature component of the three-temperature component fit given by \mbox{\citet{2010A&A...519A..34P}} has an emission measure that is comparable to the high-temperature components.
For \mbox{TW Hya}, the X-ray emission measure is dominated by a low-temperature, high density plasma at $\sim$ 3 MK which probably corresponds to an accretion shock (\mbox{\citealt{2002ApJ...567..434K}}; \mbox{\citealt{2004A&A...418..687S}}; \mbox{\citealt{2010ApJ...710.1835B}}). 
We use the three-temperature component fit given by \mbox{\citet{2006A&A...449..737R}} and we disregard the low-temperature component.
This may have a significant effect on the coronal model.
No temperature and emission measure fits are available for \mbox{CV Cha} and \mbox{GQ Lup}.

% Expanding the window makes this table very well ordered
\begin{table*}
\begin{tabular}{ccccccccccc}
\hline
 Star & 		$M_ \star$  & 			$R_ \star$ & 			$P_{rot}$ & 		Age & 		$R_{co}$ & 	$v\sin i$ & 		$\log \dot{M}_a$ & 		$\log T$ (K) & 		$\log EM$ \\
	& 		(M$_\odot$) & 		(R$_\odot$) & 		(days) & 			(Myr) & 		(R$_\star$) & 	(km s$^{-1}$) & 	(\mdot) & 				(K) & 			(cm$^{-3}$) \\ 
\hline
AA Tau & 		0.70$^1$ & 		2.00$^1$ & 		8.20$^{2,3}$ & 		1.5$^1$ & 	7.6 & 		12.3$^{4,5}$ & 		-9.2$^{1}$ & 			7.43$^{6}$ &		 52.95$^{6}$ \\
BP Tau & 		0.70$^{7}$ & 		1.95$^{8}$ & 		7.60$^{9}$ & 		1.5$^{8}$ & 	6.0 & 		9.0$^{7,8}$ & 		-8.6$^{10}$ & 			7.06$^{11}$ & 		53.28$^{11}$ \\
CR Cha & 	1.90$^{12}$ & 		2.50$^{12}$ & 		2.30$^{13}$ & 		3.0$^{12}$ & 	3.6 & 		35.0$^{14,15}$ & 	-9.0$^{12}$ & 			7.10$^{11}$ & 		53.40$^{11}$ \\
CV Cha & 	2.00$^{12}$ & 		2.50$^{12}$ & 		4.40$^{13}$ & 		5.0$^{12}$ & 	5.7 & 		25.0$^{16}$ & 		-7.5$^{12}$ & 			- & 				-\\
GQ Lup & 		1.05$^{17}$ & 		1.70$^{17}$ & 		8.40$^{17}$ & 		2.0-5.0$^{17}$ & 10.4 & 		5.0$^{17}$ & 		-9.0$^{17}$ & 			- & 				-\\
TW Hya & 	0.80$^{18}$ & 		1.10$^{18}$ & 		3.56$^{19}$ & 		8.0$^{18}$ & 	8.3 & 		5.0$^{15,20}$ & 	-8.9$^{18}$ & 			7.11$^{11}$ & 		52.64$^{11}$\\
V2129 Oph & 	1.35$^{21}$ & 		2.00$^{21}$ & 		6.53$^{22}$ & 		2.3$^{21}$ & 	8.1 &		14.5$^{15,23}$ & 	-9.2$^{21}$ & 			7.05$^{24}$ & 		53.59$^{24}$\\
V2247 Oph &	 0.36$^{10}$ & 		2.00$^{10}$ & 		3.50$^{22}$ & 		1.0$^{10}$ & 	3.4 & 		20.5$^{13}$ & 		-9.8$^{10}$ & 			7.18$^{25}$ & 		53.08$^{25}$\\
V4046 Sgr A &	 0.95$^{26}$ & 		1.12$^{26}$ & 		2.42$^{26}$ & 		15.0$^{26}$ & 	2.5 & 		13.5$^{26}$ & 		-9.3$^{26}$ & 			- & 				- \\
V4046 Sgr B &	 0.85$^{26}$ & 		1.04$^{26}$ & 		2.42$^{26}$ & 		15.0$^{26}$ & 	2.7 & 		12.5$^{26}$ & 		-9.3$^{26}$ & 			- & 				- \\
\hline
\end{tabular}
\caption{
Stellar parameters from the literature for all of the stars in the sample. 
The stars are ordered from top to bottom alphabetically. 
From left to right, the columns correspond to stellar mass, stellar radius, rotation period, age, corotation radius, projected rotational velocity, mass accretion rate, coronal temperature, and coronal X-ray emission measure. 
References for each value are given as superscripts and correspond to the following papers: 1. \mbox{\citet{2010MNRAS.409.1347D}}; 2. \mbox{\citet{1989AJ.....97..483V}}; 3. \mbox{\citet{2007A&A...463.1017B}}; 4. \mbox{\citet{1989AJ.....97..873H}}; 5. \mbox{\citet{2009ApJ...695.1648N}}; 6. \mbox{\citet{2007A&A...468..353G}}; 7. \mbox{\citet{1999ApJ...516..900J}}; 8. \mbox{\citet{2008MNRAS.386.1234D}}; 9. \mbox{\citet{1986ApJ...306..199V}}; 10. \mbox{\citet{2010MNRAS.402.1426D}}; \mbox{11. \citet{2006A&A...449..737R}}; 12. \mbox{\citet{2009MNRAS.398..189H}}; 13. \mbox{\citet{1986A&A...165..110B}}; 14. \mbox{\citet{2009ApJ...695.1648N}}; 15. \mbox{\citet{2010A&A...517A..88W}}; 16. \mbox{\citet{2003A&A...408..693S}}; 17. \mbox{\citet{2012MNRAS.425.2948D}}; 18. \mbox{\citet{2011MNRAS.417..472D}}; 19. \mbox{\citet{2008A&A...489L...9H}}; 20. \mbox{\citet{2003AJ....125..825T}}; 21. \mbox{\citet{2011MNRAS.412.2454D}}; 22. \mbox{\citet{2008A&A...479..827G}}; \mbox{23. \mbox{\citet{2007MNRAS.380.1297D}}}; \mbox{24. \mbox{\citet{2011A&A...530A...1A}}}; 25. \mbox{\citet{2010A&A...519A..34P}}; 26. \mbox{\citet{2011MNRAS.417.1747D}}.
}
\label{tbl:stellarparams}
\end{table*}

%--------------------------------------------------------------------------------------
\begin{table*}
\centering
\begin{tabular}{cccccccc}
\hline 
Star & $\Phi_{total}$ (10$^{25}$ Mx) & $<B>$ (G) & $B_{dip}$ (G) & $B_{oct}$  (G)& $\beta_{dip}$ (degrees) & $<l>$ \\
\hline
AA Tau 2009 &  	22.4 &	 1220 &	1720 & 	500 & 	170 & 	1.1 \\

BP Tau Feb 2006 & 	 15.9 & 	1010 & 	1220 & 	1600 & 	10 & 		2.3 \\

GQ Lup 2009 &		13.4 &	1170 & 	1070 & 	2430 & 	30 & 		3.0 \\

V2129 Oph 2009 & 	15.7 & 	980 & 	970 & 	2160 & 	10 & 		2.7 \\

BP Tau Dec 2006 & 	15.2 & 	970 & 	960 & 	1800 & 	30 & 		2.6 \\

GQ Lup 2011 &		10.6 & 	910 & 	900 & 	1730 & 	30 & 		2.9 \\

TW Hya 2010 & 	8.3 & 	1610 &  	730 & 	3100 & 	10 & 		2.9 \\

TW Hya 2008 & 	6.5 & 	1290 & 	370 & 	2630 & 	40 & 		3.1 \\

V2129 Oph 2005 & 	12.1 & 	740 & 	280 & 	1620 & 	20 & 		3.0 \\

CR Cha & 		6.3 &		 230 & 	220 & 	200 & 	110 & 	4.4\\

CV Cha & 		6.5 &	 	270 & 	140 & 	370 & 	120 &	 4.1\\

V2247 Oph & 		3.5 & 	220 & 	110 & 	230 & 	40 & 		4.6 \\

V4046 Sgr A & 		0.5 & 	100 & 	100 & 	130 &	 60 & 	2.5 \\

V4046 Sgr B & 		0.7 & 	150 &	 80 & 	240 & 	80 & 		3.1\\

\hline
\end{tabular}
\caption{
Results for the analysis of the ZDI magnetic maps for all of the stars in the sample.
The maps are arranged from top to bottom by decreasing strength of the dipole component of the field.
From left to right, the columns correspond to the total unsigned magnetic flux, the surface field strength averaged over the stellar surface, the strength of the dipole component, the strength of the octupole component, the tilt angle of the dipole component relative to the rotation axis, and the parameter $<l>$ defined as the energy-weighted average $l$ value that describes the field.
The parameter $<l>$ can be used as a measure of the rate at which the magnetic field strength decreases with increasing distance from the star, and can also be used as a proxy for field complexity. 
For \mbox{BP Tau}, the magnetic maps used here were produced using an experimental version of the ZDI code. However, analysis conducted with a more advanced version of the code has produced similar results (see Footnote 1 of \mbox{\citealt{2011AN....332.1027G}}).  
}
\label{tbl:CTTSmapsparameters}
\end{table*}
%----------------------------------------------------------------------------------------------------

\begin{figure*}
\includegraphics[width=0.49\textwidth]{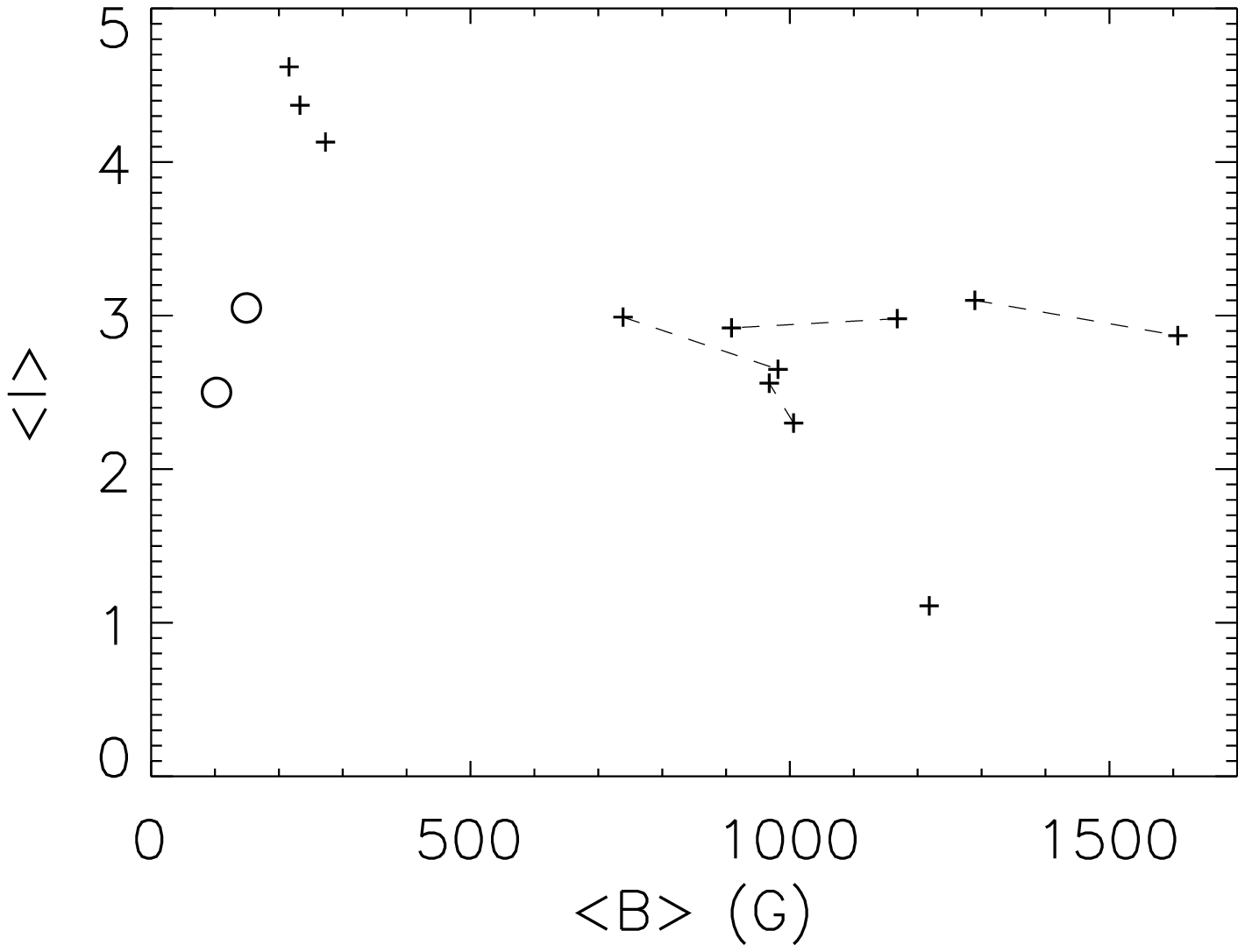}
\includegraphics[width=0.49\textwidth]{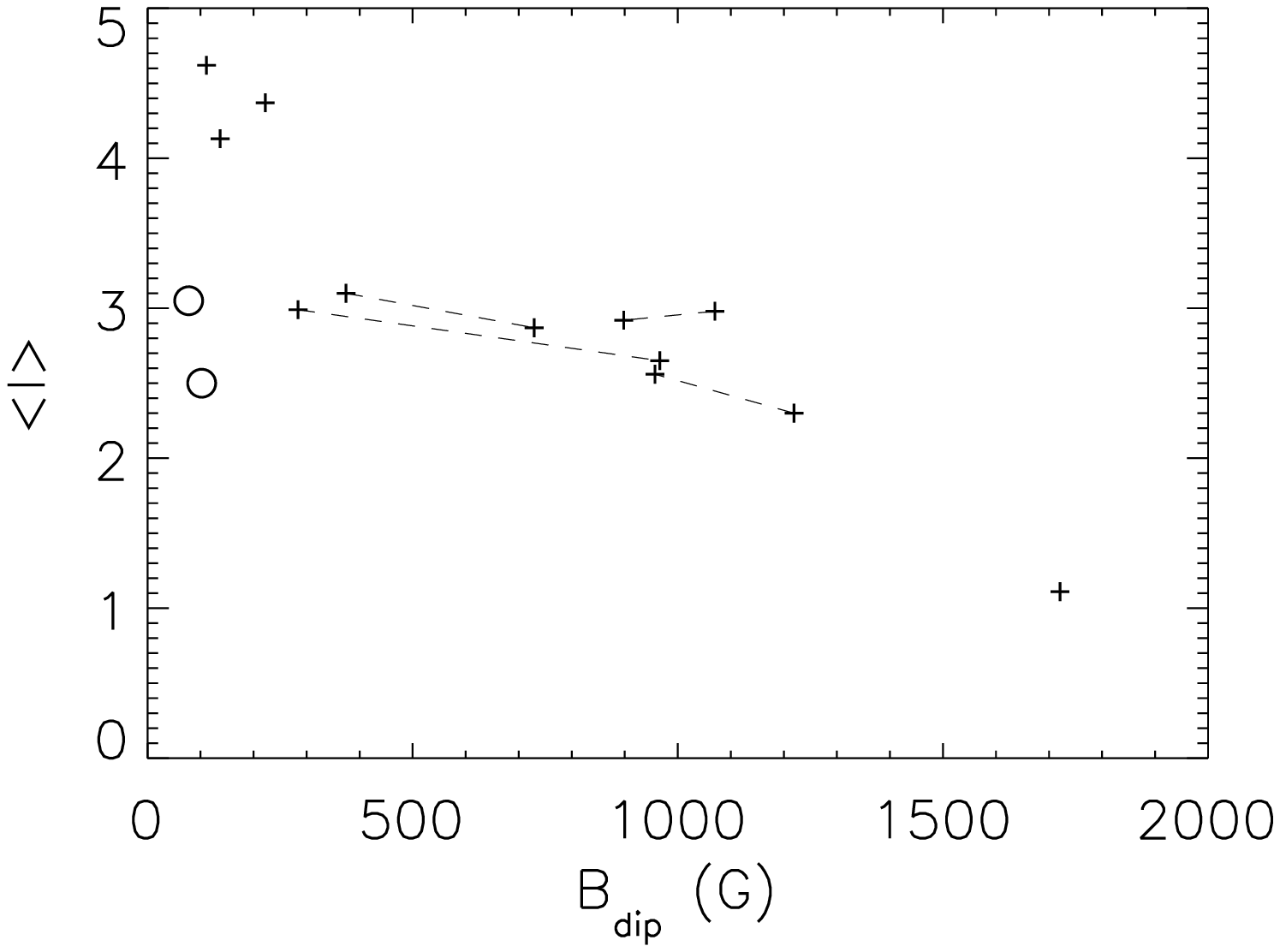}
\caption{
\emph{Left-hand panel}: correlation between $<l>$ and the surface averaged field strength $<B>$ for the sample of ZDI magnetic maps.
\emph{Right-hand panel}: correlation between $<l>$ and the strength of the dipole component of the field.  
The two stars in the binary system \mbox{V4046 Sgr} are shown as open circles. 
Data points representing the same star from two different epochs are connected by dashed lines.
}
 \label{fig:avelvsB}
\end{figure*}

\section{Surface fields} \label{sect:surfacemaps}

In this section, we analyse the available sample of ZDI maps.
Although most of these fields were fitted as the sum of poloidal and toroidal components\footnotemark, for the extrapolations that we use in this paper we assume that the fields are potential, and we therefore disregard the toroidal components of the fields.
We are therefore left with only the $B_r(\theta,\phi)$ components of the ZDI fits as these are described entirely by the poloidal components.
The $B_\theta (\theta, \phi)$ and $B_\phi (\theta, \phi)$ components are then reconstructed from $B_r (\theta,\phi)$ based on the potential field model, which we describe in more detail in Section \ref{sect:coronalmodel}.
In this model, the field is described as an expansion of spherical harmonic multipoles. 
For instance, at the stellar surface, the radial component of the field is given by

\begin{equation} \label{eqn:Brsurface}
B_r (\theta,\phi) = \sum_{l=1}^{\infty} \sum_{m=-l}^{l} c_{lm} P_{lm} (\cos \theta) e^{im\phi}
\end{equation} 

\footnotetext{
It is common in the literature to define the poloidal component of a magnetic field as a combination of the $B_r$ and $B_\theta$ components, and the toroidal component as the $B_\phi$ components. 
Within the ZDI community, the definitions of these terms are different and are described in Appendix III of \mbox{\citet{1961hhs..book.....C}}.
}

\noindent where $P_{lm} (\cos \theta)$ are the associated Legendre polynomials, $\theta$ is the colatitude (with values of 0 and $\pi$ at the poles of the coordinate system), $\phi$ is the longitude, and $c_{lm}$ characterises the strengths of the spherical harmonic components.
In this model, the $l = 1$ components represent dipole fields, the $l=2$ components represent quadrupole fields, and the \mbox{$l=3$} components represent octupole fields.
In general, higher \mbox{$l$-values} represent more complex fields.

It is useful to have a simple way to quantify how magnetic energy is distributed between high- and low-order spherical harmonic components\footnotemark.
For this, we use the energy-weighted average $l$ value that describes the field. 
This is given by

\footnotetext{
We define the magnetic energy as the magnetic energy density, $B^2$, integrated over the stellar surface. 
Although this does not actually give an energy, if we integrate this outwards in radius by a small distance, the magnetic field can be assumed to be uniform over this distance.
Since we only ever consider the energy in each spherical harmonic component as the fraction of the total energy in the field, this integral outwards in radius can be ignored. 
}

\begin{equation}
<l> = \frac{\sum_{l=1}^{\infty}  E_l l}{\sum_{l=1}^{\infty} E_l}
\end{equation}

\noindent where $E_l $ is the magnetic energy held within each \mbox{$l$-value} component. 
We compute this quantity using only the poloidal component of the field so that the results can be compared to the field extrapolations in the following sections. 
Since the strengths of higher $l$-value components decreases with distance from the star faster than the strengths of lower $l$-value components, this is a simple measure of the rate at which the star's magnetic field strength falls off with distance from the star (see Section \ref{sect:corona}).
This can also be thought of as a simple measure of the complexity of a magnetic field because fields with simple dipolar configurations have low values of $<l>$, and fields with lots of small-scale features have higher values of $<l>$.
However, this does not take into account the amount of magnetic energy in non-axisymmetric spherical harmonic components, which can be considered when constructing a more representative measure of field complexity.

\subsection{Results}

The radial components of the previously published magnetic maps for \mbox{AA Tau} and \mbox{V2247 Oph}, and the new maps for \mbox{CR Cha} and \mbox{CV Cha} are shown in \mbox{Fig. \ref{fig:CTTSdispAll}}.
There is significant variation in magnetic field complexity across all of the magnetograms in the sample.  
Some stars have large-scale fields that are well described with dipole plus octupole components, namely, in increasing/decreasing importance of the octupole/dipole component, \mbox{AA Tau}, \mbox{BP Tau}, \mbox{GQ Lup}, \mbox{V2129 Oph}, and \mbox{TW Hya}.  
The other stars, \mbox{V2247 Oph}, \mbox{CR Cha}, \mbox{CV Cha}, \mbox{V4046 Sgr A}, and \mbox{V4046 Sgr B}, have more complex field topologies.
In \mbox{Table \ref{tbl:CTTSmapsparameters}}, we give the total unsigned magnetic fluxes, surface averaged field strengths, $<l>$-values, and the strengths of the dipole and octupole components for each of the fields.
We define the strength of the dipole component as the polar field strength of the dipole component (i.e. the polar field strength of a field that consists of the dipole component only), and we use a similar definition for the strength of the octupole component.
The strengths of the dipole components are important as these components dominate far from the star and therefore primarily determine the open magnetic fluxes and disc truncation radii.
The octupole components are important because ZDI studies suggest that CTTSs often possess very strong octupole components. 
As we do not consider the toroidal components of the fields, the surface averaged field strengths are underestimates for the true values derived from the ZDI fits.

The simplest large-scale field in the sample is on the fully convective star \mbox{AA Tau}, which has a strong dipole component of strength 1.7 kG, and an octupole component of strength 500 G.
The field has an $<l>$-value of 1.1, which means that almost all of the magnetic energy is held in the dipole component of the field.  
The stars \mbox{BP Tau}, \mbox{GQ Lup}, \mbox{TW Hya}, \mbox{V2129 Oph}, \mbox{V4046 Sgr A}, and \mbox{V4046 Sgr B} also possess simple large-scale fields, with $<l>$-values of $\sim$ 2-3. 
In each case, the octupole component has a much stronger polar field strength than the dipole component.
For the two \mbox{BP Tau} fields, and the \mbox{V2129 Oph 2009} field, the dipole and octupole components have equal strengths at low-latitudes.
For the two \mbox{TW Hya} fields, and the V2129 Oph 2005 field, the octupole component dominates at all latitudes.
Due to the strong $\sim$ 3 kG octupole component on \mbox{TW Hya}, this field has the largest value of the surface averaged field strength.
Due to its relatively old age, \mbox{TW Hya} has a relatively small surface area; this means that it has a much smaller unsigned magnetic flux.
All of these fields have surface averaged field strengths above 500 G.

 \begin{figure}
\includegraphics[width=0.49\textwidth]{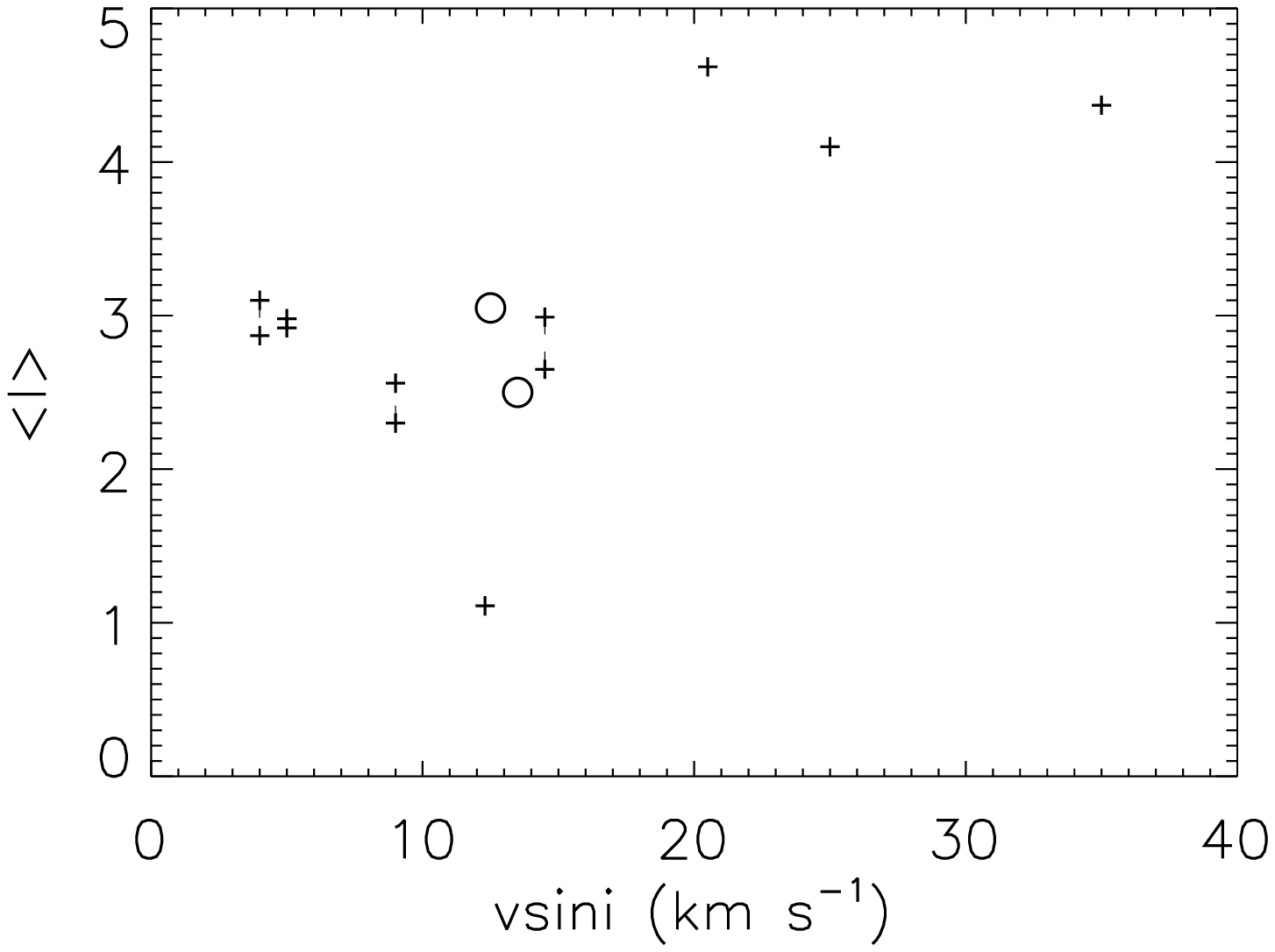}
\includegraphics[width=0.49\textwidth]{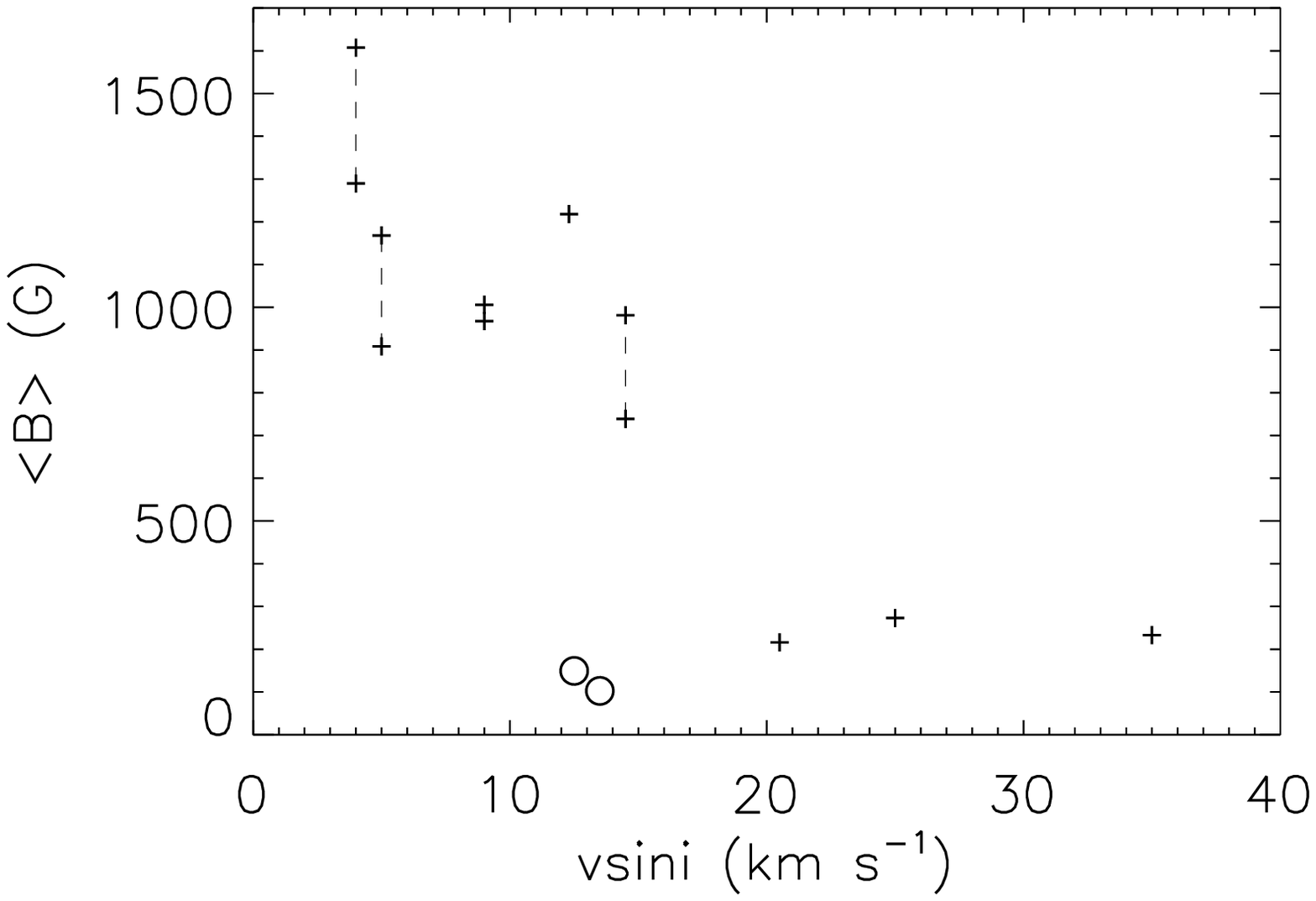}
\caption{
\emph{Upper panel}: correlation between the projected rotational velocity and $<l>$.
\emph{Lower panel}: correlation between the projected rotational velocity and the surface averaged field strength. 
The two stars in the binary system \mbox{V4046 Sgr} are shown as open circles. 
Data points representing the same star from two different epochs are connected by dashed lines.
}
\label{fig:mapparamsvsini}
\end{figure}

The magnetic fields on \mbox{CR Cha}, \mbox{CV Cha}, and \mbox{V2247 Oph}, have larger $<l>$-values between 4-5. 
For \mbox{V2247 Oph}, the polar regions show single polarity fields, and the low-latitude fields are covered in complex patterns of regions of opposite polarity field. 
For \mbox{CR Cha} and \mbox{CV Cha}, the complex patterns of regions of opposite polarity field cover the stellar surfaces at all latitudes. 
All of these stars have very weak large-scale fields, with surface averaged field strengths of a few hundred Gauss.

In \mbox{Fig. \ref{fig:avelvsB}}, we show that there is some correlation between $<l>$ and $<B>$.
The simple fields are the strongest, and the more complex fields are the weakest. 
The possible exceptions are the fields on \mbox{V4046 Sgr A} and \mbox{V4046 Sgr B}, which appear to have relatively weak fields given their small values of $<l>$. 
A similar correlation between field strength and field complexity is seen on the M dwarfs that have been studied using ZDI (\mbox{\citealt{2008MNRAS.390..545D}}; \mbox{\citealt{2008MNRAS.390..567M}}; \mbox{\citealt{2010MNRAS.407.2269M}}).

Since the spatial resolution of the ZDI technique is dependent on stellar rotation rate - the fastest rotators have the highest resolution - it is important to understand to what extent the differences in the complexities of the ZDI maps are resolution effects.
\mbox{\citet{2009MNRAS.398..189H}} explored this possibility for \mbox{CR Cha}, \mbox{CV Cha}, \mbox{BP Tau}, and \mbox{V2129 Oph}. 
These stars have projected rotational velocities of \mbox{35 km s$^{-1}$}, \mbox{25 km s$^{-1}$}, \mbox{9 km s$^{-1}$}, and \mbox{14.5 km s$^{-1}$} respectively.
They produced simulated data for \mbox{CR Cha} and \mbox{CV Cha} based on the reconstructed magnetic maps for the two stars, and assuming the stellar parameters of \mbox{BP Tau} and \mbox{V2129 Oph}.
Using this simulated data, they again reconstructed the magnetic topologies for \mbox{CR Cha} and \mbox{CV Cha}. 
They found that the stellar parameters of \mbox{V2129 Oph} gave magnetic maps for \mbox{CR Cha} and \mbox{CV Cha} of similar complexity to the magnetic map of \mbox{V2129 Oph} \emph{but with much smaller field strengths}. 
However, when they assumed the \mbox{BP Tau} parameters, their reconstructed fields were significantly more complex that the field found on \mbox{BP Tau}. 
This indicates that the large-scale magnetic field of \mbox{BP Tau} is genuinely much simpler than those of \mbox{CR Cha}, \mbox{CV Cha}, and \mbox{V2129 Oph}; if \mbox{BP Tau} hosted a more complex large-scale field, it would have been detected by ZDI.

In \mbox{Fig. \ref{fig:mapparamsvsini}}, we show a clear correlation between $<l>$ and projected rotational velocity. 
Although this is consistent with the interpretation that the differences between the complexities of the magnetic maps is a resolution effect since greater field complexity can only be recovered on faster rotators, if this interpretation was correct, we would expect that the strongest field strengths would be seen on the rapid rotators due to a lower amount of flux cancellation.
However, these stars show the opposite trend, as we show in \mbox{Fig. \ref{fig:mapparamsvsini}}.
Also, in the sample of M dwarfs with known large-scale fields geometries, reconstructed using the same ZDI technique, there is no clear correlation between field complexity and rotation rate.
In fact, in many cases, the rapidly rotating main-sequence M dwarfs have the simplest fields.

\begin{figure}
\includegraphics[width=0.49\textwidth]{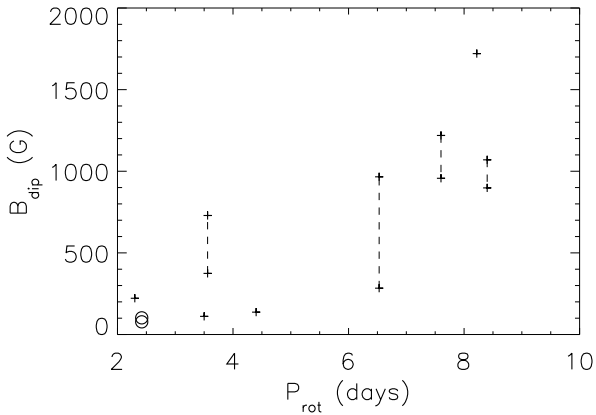}
\caption{
Correlation between the strength of the dipole component of the field and the stellar rotation period. 
Stars with weak dipole components tend to be rotating faster than stars with strong dipole components.
The two stars in the binary system \mbox{V4046 Sgr} are shown as open circles. 
Data points representing the same star from two different epochs are connected by dashed lines.
}
\label{fig:BdipvsP}
\end{figure}

The correlation between $<l>$, field strength, and rotation rate may be a result of star-disc magnetic interactions. 
Coupling between the stellar magnetic field and the disc leads to torques being exerted on the central star.
Magnetic field lines extending from the star to regions of the disc inside the corotation radius exert a spin-up torque on the star; magnetic field lines extending from the star to regions of the disc beyond the corotation radius exert a spin-down torque on the star.
Stars with stronger dipole fields might be expected to rotate slower for two reasons.
Firstly, a stronger dipole field will truncate the inner edge of the disc further from the stellar surface, which might lead to a smaller spin-up torque on the central star.
At the same time, a stronger dipole field will couple to the disc beyond the corotation radius more efficiently, leading to a greater spin-down torque. 
Thus, the correlation between rotation rate and the strength of the dipole component, seen in \mbox{Fig. \ref{fig:BdipvsP}}, could be a natural consequence of star-disc interactions. 
As the strength of the dipole component is positively correlated with the field complexity and the average field strength, this can explain why the faster rotators have stronger and more complex fields.

\begin{figure*}
\centering
\subfigure[AA Tau 2009]{\includegraphics[width=0.32\textwidth]{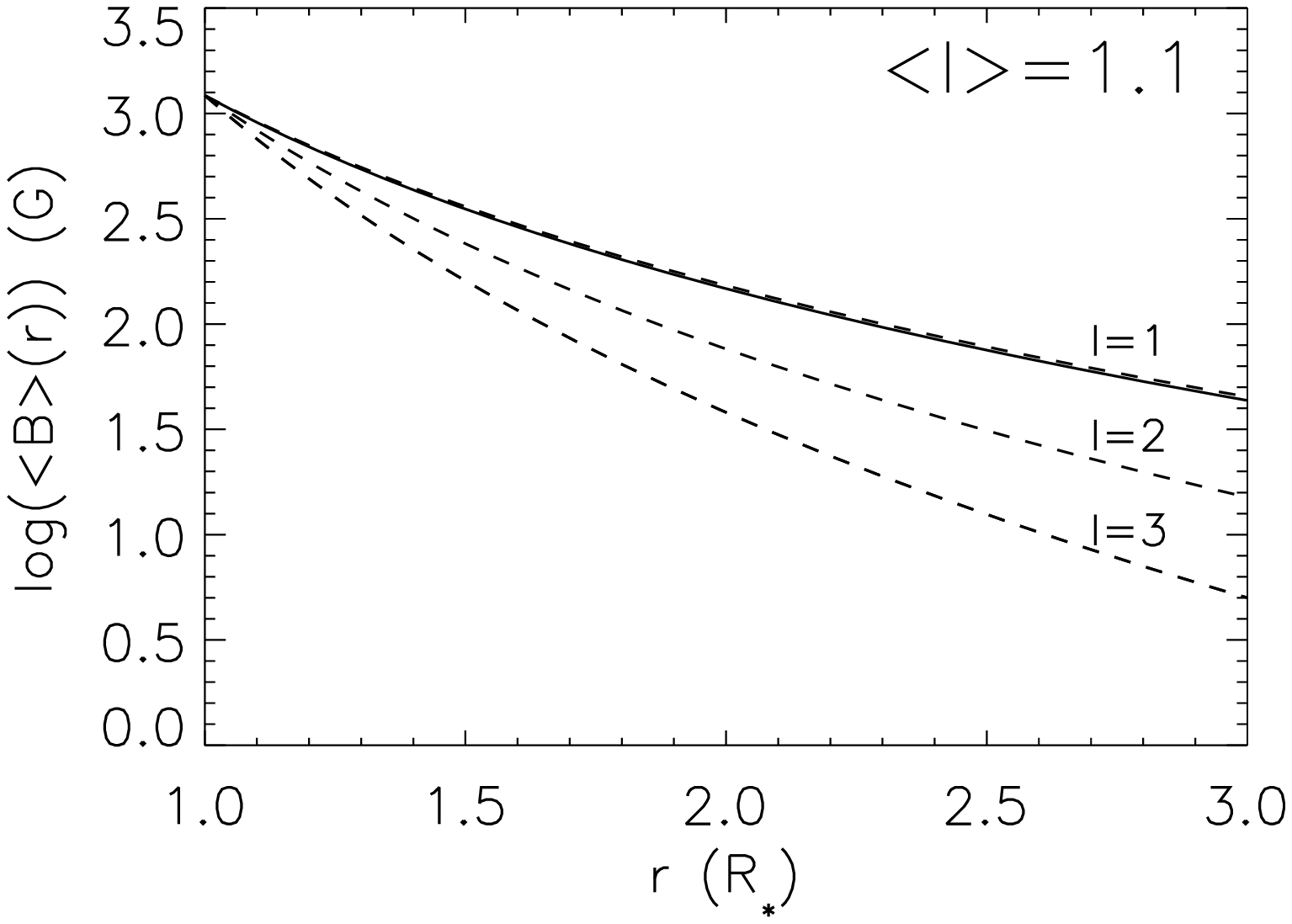}}
\subfigure[TW Hya 2010]{\includegraphics[width=0.32\textwidth]{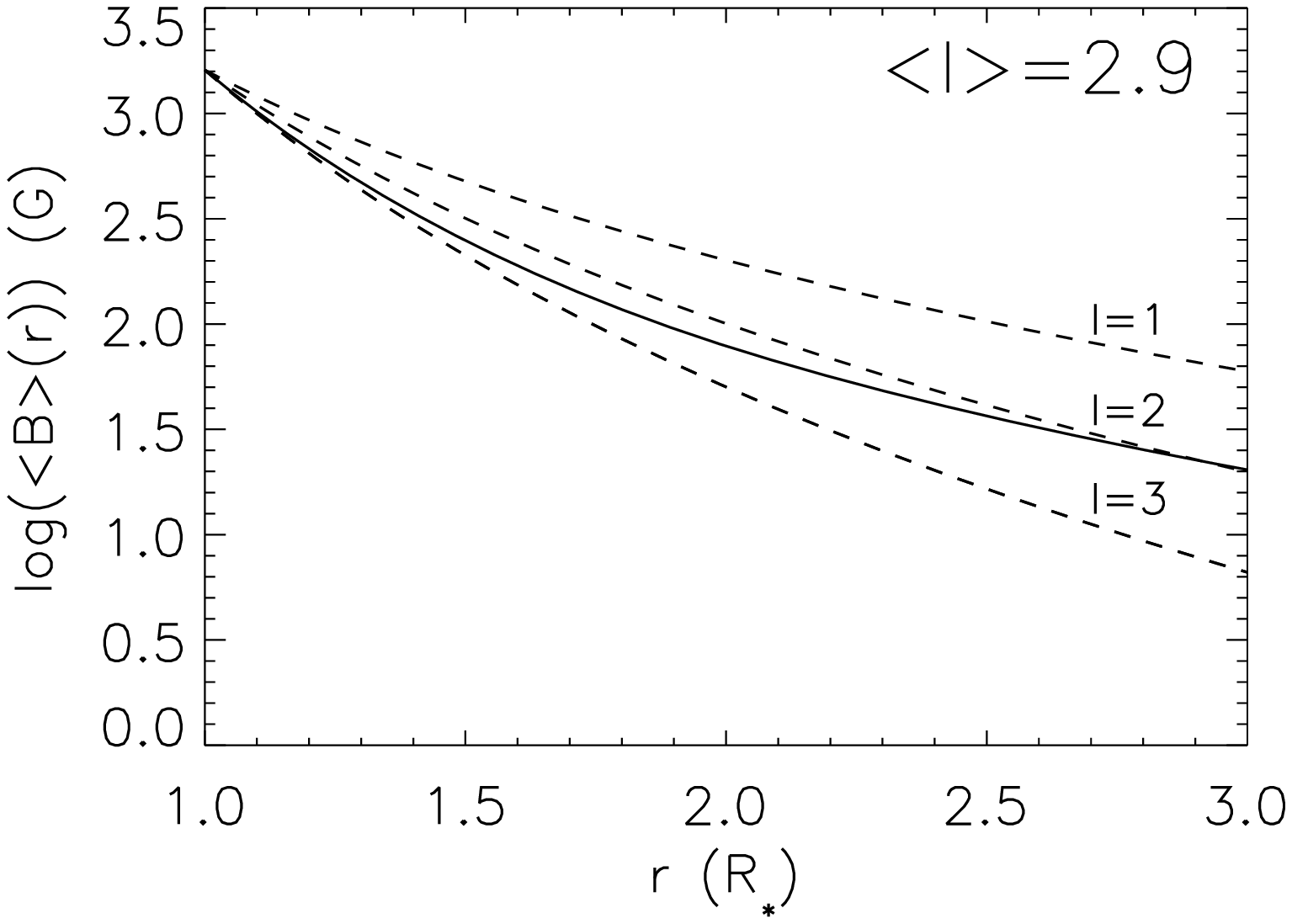}}
\subfigure[V2247 Oph]{\includegraphics[width=0.32\textwidth]{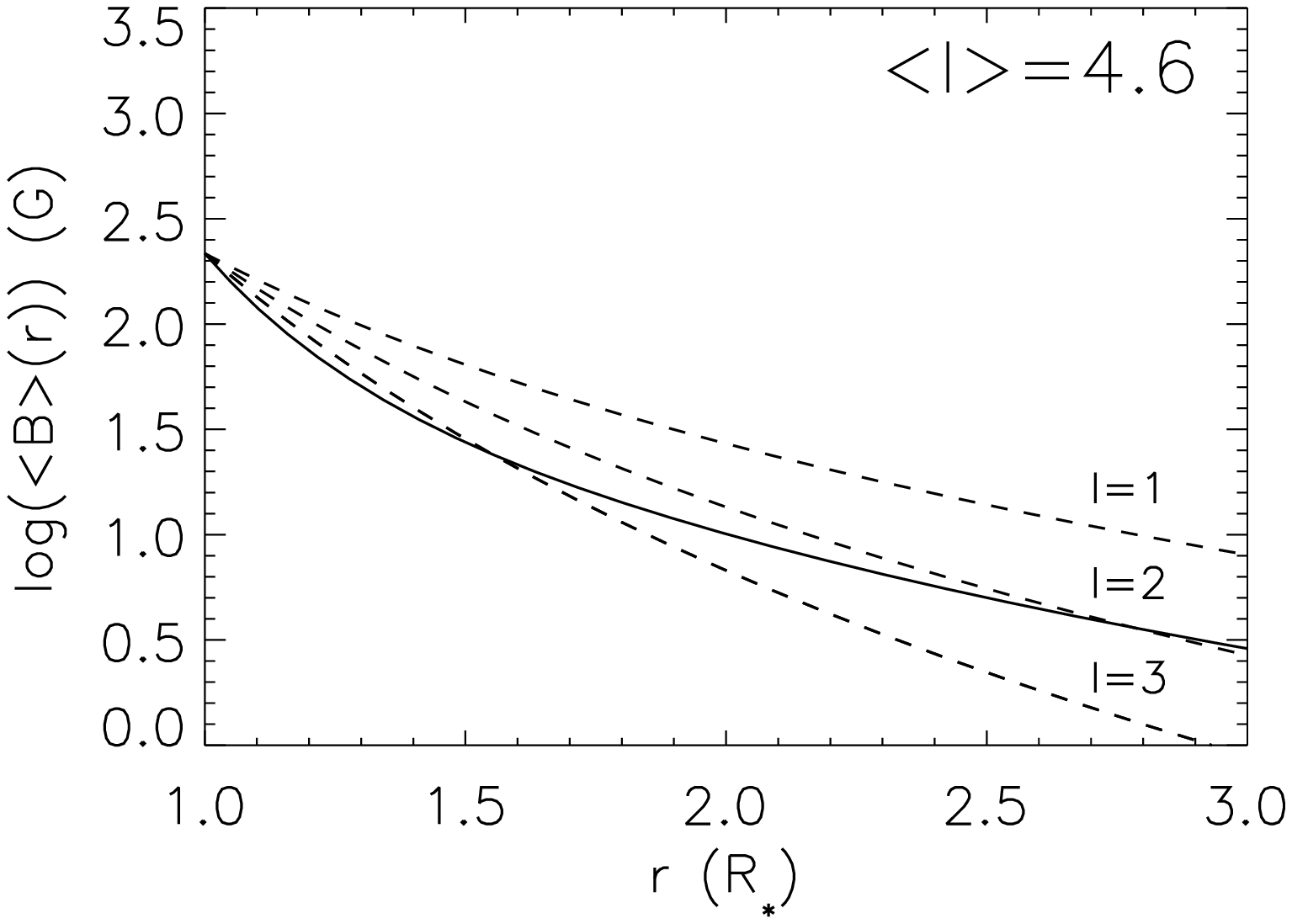}}
\caption{
Figures showing magnetic field strengths, averaged over all latitude and longitude, as a function of distance from the star for the field extrapolations (with no source-surface assumed) of \mbox{AA Tau}, \mbox{TW Hya 2010}, and \mbox{V2247 Oph}.
The magnetic fields of \mbox{AA Tau}, \mbox{TW Hya 2010}, and \mbox{V2247 Oph} are good examples of dipole fields, octupole fields, and more complex fields respectively, and the 3D potential field extrapolations of these magnetic fields can be seen in \mbox{Fig. \ref{fig:acclines}}.  
The corresponding field strengths, as functions of radius, for dipole ($l=1$), quadrupole ($l=2$), and octupole ($l=3$) fields with the same surface field strengths are shown as dashed lines.
This shows how fast the strengths of magnetic fields of difference complexities decrease with increasing radius.
The decrease in the steepness of the slopes with increasing radius for both \mbox{TW Hya 2010} and \mbox{V2247 Oph}, shows how the influence of high-order spherical harmonic components diminishes far from the stellar surface. 
Close to the stellar surfaces of \mbox{TW Hya} and \mbox{V2247 Oph}, the field strengths decrease with increasing radius faster than for the more dipolar field of \mbox{AA Tau}.
However, further from the star, the field strengths decrease with radius at the same rate as the \mbox{AA Tau} field. 
Despite the different field complexities, in all three cases, the field strength falls off far from the star at the same rate as a dipole ($B \propto r^{-3}$).
This is due to the dominance of the dipole component far from the stellar surface.
}
 \label{fig:ratedropoff}
\end{figure*}

\section{Closed X-ray emitting coronae} \label{sect:corona}

\subsection{Model} \label{sect:coronalmodel}

\begin{figure*}
\centering
\includegraphics[width=0.49\textwidth]{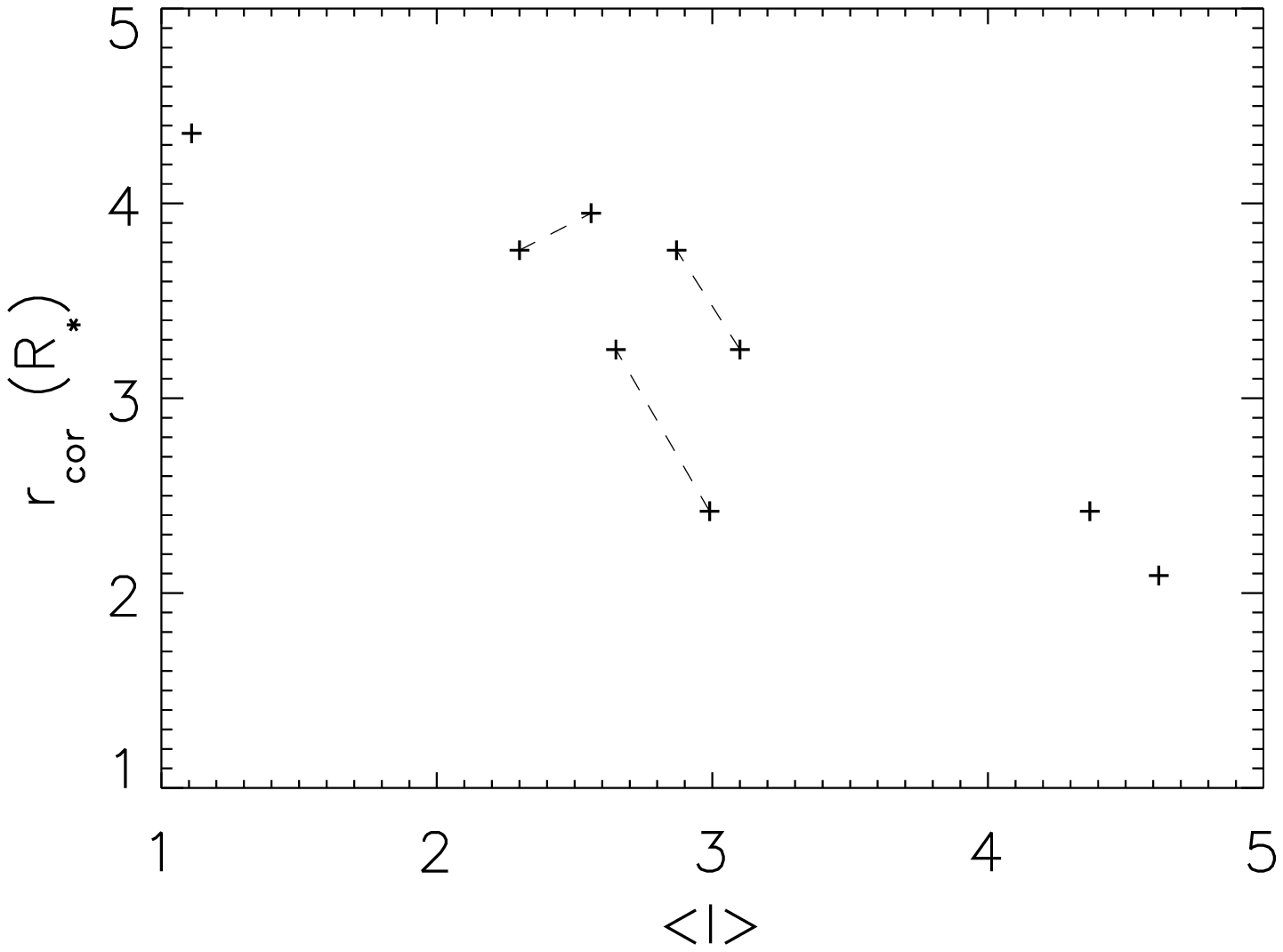}
\includegraphics[width=0.49\textwidth]{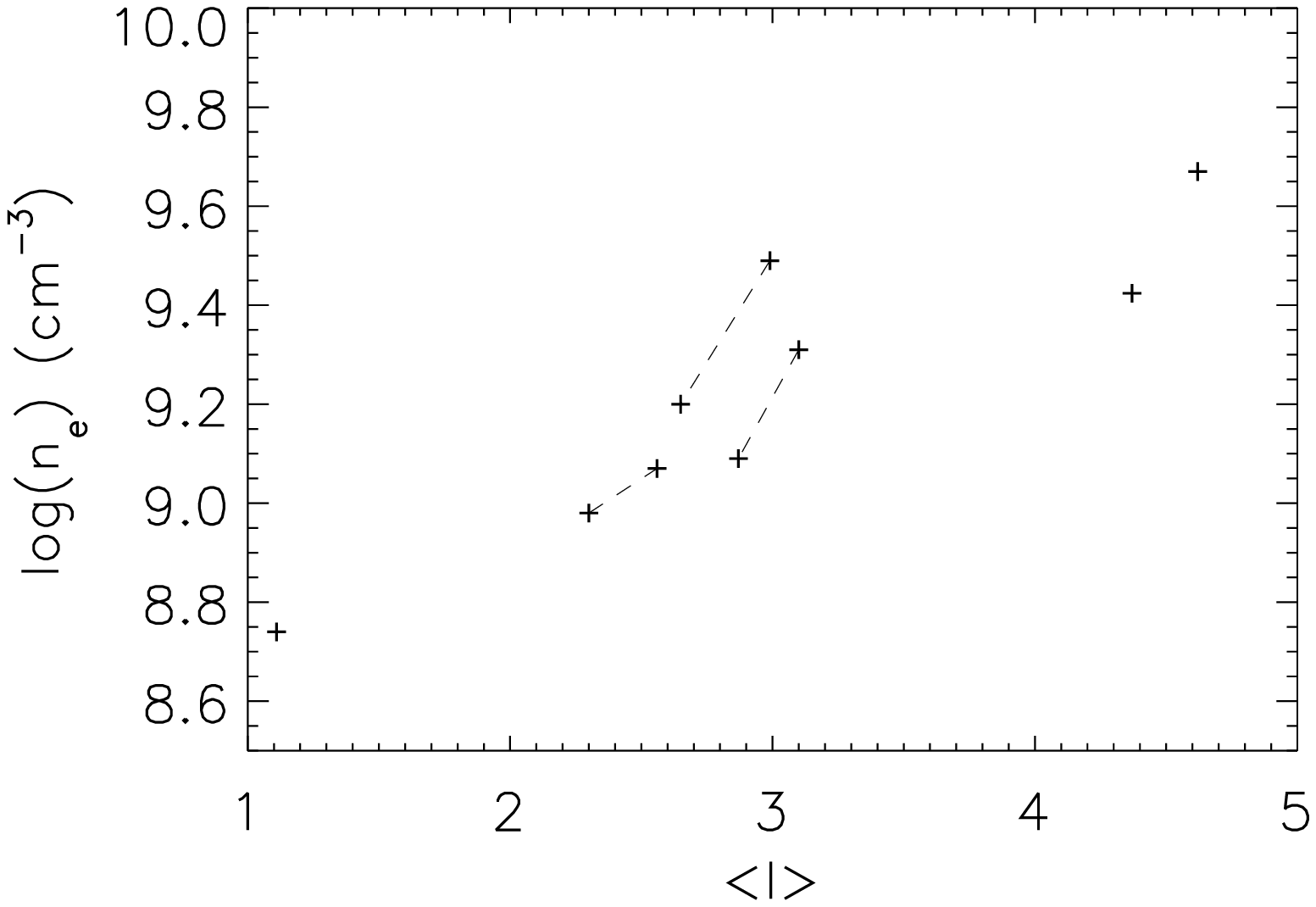}
\caption{
Results of the coronal X-ray emitting plasma model. 
\emph{Left-hand panel}: the maximum radius to which the stellar magnetic fields can contain the hot coronal plasma as a function of $<l>$. 
\emph{Right-hand panel}: the emission measure weighted average coronal electron density as a function of $<l>$. 
These figures show that simple magnetic fields lead to extended, diffuse coronae, and complex fields lead to compact, dense coronae. 
Data points representing the same star from two different epochs are connected by dashed lines.
}
 \label{fig:extentcomplexity}
\end{figure*}

\begin{table*}
\begin{tabular}{ccccccccc}
\hline
 Star/Magnetograms & $\log K $ & $\log \bar{n}_e$ (cm$^{-3}$) & $r_{cor}$ (R$_\star$) & V ($10^{35}$ cm$^3$) & $\Phi_{open} (10^{25}$ Mx$)$ & $f_{open}$ & $< |B_{r,open}|>$ (G) \\
\hline

AA Tau 2009 & 	-6.7 & 	8.7 & 	4.4 & 	3.7 & 	5.3 & 	0.17 & 	1330\\

BP Tau Feb 2006 & 	-6.9 & 	9.0 & 	3.8 & 	2.0 & 	4.3 & 	0.11 & 	1830 \\

V2129 Oph 2009 & 	-6.9 & 	9.2 & 	3.3 & 	1.3 & 	4.1& 		0.06 & 	2810\\

BP Tau Dec 2006 & 	-6.8 & 	9.1 & 	4.0 & 	1.8 & 	4.0 & 	0.16 & 	960\\

TW Hya 2010 & 	-6.7 & 	9.1 & 	3.8 & 	0.3 & 	1.0 & 	0.09 & 	1540\\

TW Hya 2008 & 	-6.6 & 	9.3 & 	3.3 & 	0.2 & 	0.7 & 	0.06 & 	1510\\  

V2129 Oph 2005 & 	-6.0 & 	9.5 & 	2.4 & 	0.4 & 	2.0 & 	0.08 & 	1120\\

CR Cha  & 		-5.2 & 	9.4 & 	2.4 & 	0.7 & 	6.3 & 	0.33 & 	210\\

V2247 Oph & 		-5.0 & 	9.7 & 	2.1 & 	0.1 & 	1.1 & 	0.20 & 	230\\

\hline
\end{tabular}
\caption{
Results from the coronal plasma model for each of the stars in the sample.
The maps are arranged from top to bottom by descending strength of the dipole component of the field.
From left to right, the columns correspond to the proportionality parameter $K$ between the magnetic pressure and the plasma pressure at the base of the corona fitted to observed coronal emission measures, the emission measure weighted average coronal electron density, the maximum radius at which the stellar magnetic field can contain the hot coronal plasma, the volume of the closed corona, the open unsigned magnetic flux, the fraction of the stellar surface covered in open field, and the average field strength in regions of the stellar surface covered in open field.
}
\label{tbl:cttscoronaresults}
\end{table*}

Based on the solar analogy, we model the coronal magnetic fields of the CTTSs as potential-fields (meaning that they contain no electric currents) using the potential-field source-surface (PFSS) model. 
Although stellar closed coronae contain hot plasma that alters the magnetic field structures, in the Sun, the PFSS model is reasonably accurate when describing large-scale magnetic field geometries.
This has been shown by comparisons of closed coronal structures predicted with the PFSS model with images of the green line solar corona (\mbox{\citealt{1997ApJ...485..419W}}) and by comparisons of the PFSS model with the results of MHD modelling for the solar corona \mbox{(\citealt{2006ApJ...653.1510R})} and for other stars (\mbox{\citealt{2013MNRAS.431..528J}}). 
Although in the large scale corona, we do not expect currents to significantly alter the magnetic field structure, on small-scales, the field structures will be significantly altered by sub-surface convective motions.
Since we do not model small-scale field structures, this issue is not considered in this paper.
It is unclear to what extent large stellar flares affect coronal magnetic field structures, and such effects are not included in our models. 
Our models therefore correspond to quiescent (non-flaring) coronae.

Assuming that the magnetic field is potential (current-free) means that it can be described entirely by the gradient of a magnetostatic scalar potential $\Psi$, where $\mathbf{B} = -\nabla \Psi$.
In this case, $\Psi$ must satisfy Laplace's equation $\nabla^2 \Psi = 0$, the solution of which can be written as a spherical harmonic expansion:

\begin{equation} \label{eqn:Laplacesolution}
\Psi (r,\theta,\phi) = \sum_{l=1}^{\infty} \sum_{m=-l}^{l} \left[ a_{lm} r^l + b_{lm} r^{-(l+1)} \right] P_{lm} (\cos \theta) e^{im\phi}.
\end{equation}

\noindent If the field strength is assumed to vanish very far from the star, then the values of $a_{lm}$ must all be zero. 
However, the coronal magnetic field structures that are consistent with this assumption are unrealistic because they do not take into account the opening of magnetic field lines by coronal plasma.
This happens approximately at the point where the plasma $\beta$ becomes greater than unity, and the field lines beyond that point are approximately radial.
This can be corrected for in the potential-field model by assuming a spherical source-surface with radius $R_S$ where the field becomes entirely radial, i.e. \mbox{$B_\theta (R_S) = B_\phi (R_S) = 0$}.
Making the source-surface assumption, and defining \mbox{$c_{lm} = b_{lm} (l+1) R_\star^{-(l+2)} - a_{lm} l R_\star^{l-1}$}, gives

\begin{equation} \label{eqn:Br2}
B_r(r,\theta,\phi) = \sum_{l=1}^{\infty} \sum_{m=-l}^{l} c_{lm} f_{l}(r) P_{lm} (\cos \theta) e^{i m \phi}
\end{equation}

\begin{equation} \label{eqn:Bt2}
B_\theta(r,\theta,\phi) = \sum_{l=1}^{\infty} \sum_{m=-l}^{l}  c_{lm}  g_{l}(r) \frac{d P_{lm}(\cos \theta)}{d \theta}  e^{i m \phi}
\end{equation}

\begin{equation} \label{eqn:Bp2}
B_\phi(r,\theta,\phi) = \sum_{l=1}^{\infty} \sum_{m=-l}^{l}  c_{lm} g_{l}(r) P_{lm} (\cos \theta) \frac{im}{\sin \theta} e^{i m \phi}
\end{equation}

\noindent where $f_{l}(r)$ and $g_{l}(r)$ are given by

\begin{equation}
f_{l}(r) = \left[ \frac{(l+1)\left(\frac{r}{R_\star} \right)^{-(l+2)} + l \left(\frac{R_S}{R_\star} \right)^{-(2l+1)} \left( \frac{r}{R_\star} \right)^{l-1}  }{l \left( \frac{R_S}{R_\star} \right)^{-(2l+1)} + l + 1} \right] 
\end{equation}

\begin{equation}
g_{l}(r) = \left[ \frac{ \left( \frac{r}{R_\star} \right)^{l-1}  \left( \frac{R_S}{R_\star} \right)^{-(2l+1)} - \left( \frac{r}{R_\star} \right)^{-(l+2)}}{l \left( \frac{R_S}{R_\star} \right)^{-(2l+1)} + l + 1} \right].
\end{equation}

\noindent This is the potential-field source-surface model (PFSS), developed by \mbox{\citet{1969SoPh....9..131A}} and \mbox{\citet{1969SoPh....6..442S}}, and is discussed in detail by \mbox{\citet{2010RPPh...73l6901G}} and \mbox{\citet{2012PhDT........75J}}.
When the source-surface radius is very large, this is equivalent to taking $a_{lm} = 0$ in \mbox{Eqn. \ref{eqn:Laplacesolution}}.
At the stellar surface, $f_l (r) = 1$, and \mbox{Eqn. \ref{eqn:Br2}} reduces to \mbox{Eqn. \ref{eqn:Brsurface}}.
The values of the $c_{lm}$ coefficients are calculated using only the radial components of the ZDI maps.
This means that the non-potential components of the ZDI maps are not taken into account in the extrapolations.  
\mbox{\citet{2013MNRAS.431..528J}} used MHD stellar wind models to investigate the effects that ignoring toroidal fields can have on field extrapolations and found that the presence of toroidal surface fields is incompatible with a wind solution that is in force balance, and therefore such fields must be confined close to the stellar surface, below the heights at which the wind is launched.
The code that we use to extrapolate the fields was originally developed by \mbox{\citet{1998ApJ...501..866V}}.
The free parameter in this model is the source-surface radius, $R_S$, which we set to the equatorial corotation radius.
At this radius, the exact choice of the source-surface radius is unlikely to have a significant effect on the coronal plasma model (\mbox{\citealt{2008MNRAS.386..688J}}).

Using the field extrapolations, we model the closed coronal plasma by assuming an isothermal corona in hydrostatic equilibrium that is corotating with the stellar surface.
This model has been developed by \mbox{\citet{2002MNRAS.336.1364J,2006MNRAS.367..917J,2008MNRAS.386..688J}}.
The isothermal and hydrostatic equilibrium assumptions mean that the plasma pressure along each magnetic loop is given by 

\begin{equation} \label{eqn:hydrostaticequib}
p = p_0 \exp \left(  \frac{\bar{m}}{kT} \int g_s ds \right)
\end{equation}

\noindent where $p_0$ is the plasma pressure at the base of each field line, $\bar{m}$ is the mean molecular mass, $s$ is the coordinate measured along the path of the magnetic loop, and $g_s$ is the component of the effective gravity parallel to the field line.
This is given by 

\begin{equation}
g_s = \frac{\bf{g} . \bf{B}}{|\bf{B}|} 
\end{equation}

\noindent where 

\begin{equation}
\mathbf{g}(r,\theta) = \left( - G M_\star / r^2 + \omega^2 r \sin^2 \theta, \omega^2 r \sin \theta \cos \theta \right).
\end{equation}

\noindent Here $M_\star$ is the stellar mass, $r$ is the distance from the centre of the star, $\omega$ is the stellar surface angular velocity (we ignore differential rotation), and $\theta$ is the colatitude coordinate. 
Although the opening of coronal field lines is simulated using the source-surface assumption in the field extrapolation model, we further apply the condition that the density is zero at any point if the field line through that point has a gas pressure that exceeds the magnetic pressure anywhere along its length. 
Such field lines are assumed to be dark in X-rays and do not contribute to the X-ray emission measure.

We assume that  the plasma pressure at the base of the corona, $p_0$, is proportional to the magnetic pressure, such that

\begin{equation} \label{eqn:pressurescaling}
p_0 = K B_0^2
\end{equation}

\noindent where $B_0$ is the average of the magnetic field strengths at the two footpoints of the magnetic loop. 
This ensures that regions of strong magnetic field give the highest densities and is similar to the model of \mbox{\citet{1997ApJ...485..419W}}.
They modelled the solar corona using field extrapolations of synodic magnetograms and fitted their model to simultaneous green light observations of the corona.
We assume that $K$ is uniform over the surface of each star, but varies from star to star.
In order for there to be a closed corona, the value of $K$ must be significantly less than unity.
For each star, we scale $K$ such that the modelled global emission measures (i.e. $n_e^2$ integrated over the closed coronal volume) fit observed X-ray emission measures from the literature.
The coronal temperatures and emission measures compiled from the literature for each star are discussed in Section \ref{sect:sample}. 
As no coronal temperatures and emission measures are available for \mbox{CV Cha} and \mbox{GQ Lup} in the literature, we do not consider them in this section. 
We also do not consider the two stars in the close binary system \mbox{V4046 Sgr} as the nature of the system is much more complex than the others in the sample.

Due to the reliance of the ZDI technique on Stokes V (circularly polarised) line profiles, ZDI magnetic maps suffer from missing magnetic flux in small-scale field structures.
\mbox{\citet{2011ASPC..448..255R}} compared surface magnetic fluxes measured using Stokes I and Stokes V for a number of M dwarf main-sequence stars, and found that Stokes V typically reproduced only $\sim$5\% of the magnetic flux found using Stokes I.
This missing flux is probably contained within small-scale active regions similar to those found on the solar surface. 
\mbox{\citet{2010MNRAS.404..101J}} and \mbox{\citet{2011MNRAS.410.2472A}} investigated the effects of missing magnetic flux on the coronal models used in this paper. 
They found that when magnetic maps are dominated by small-scale field structures, the coronal X-ray emission is dominated by these field structures.
Therefore, the possibility that large amounts of magnetic flux is missing on small-scales in the ZDI magnetic maps represents a limitation to our models. 
Nevertheless, the magnetic maps that we use in this paper represent the highest quality data available from current instrumentation. 
In \mbox{Appendix \ref{appendix:smallscalefield}}, we test the effects of missing small-scale field structures on the results presented in this section.
The loss of small scale field structures is important mostly due to the effect on the $K$ parameter in the model, which we fit to observationally constrained emission measures.   
When magnetic flux in the magnetic map is lost, a larger value of $K$ is needed to reproduce the observed emission measure, which corresponds to an increase in the plasma-$\beta$ at the base of the corona. 
We find that our calculated open fluxes are not affected by the loss of small-scale field structures, which is consistent with the result in the following section that the open flux is determined by the strength of the dipole component of the field.
We also find moderate decreases in our the closed coronal extents, and increases in the open flux filling factors.

\begin{figure}
\centering
\includegraphics[width=0.49\textwidth]{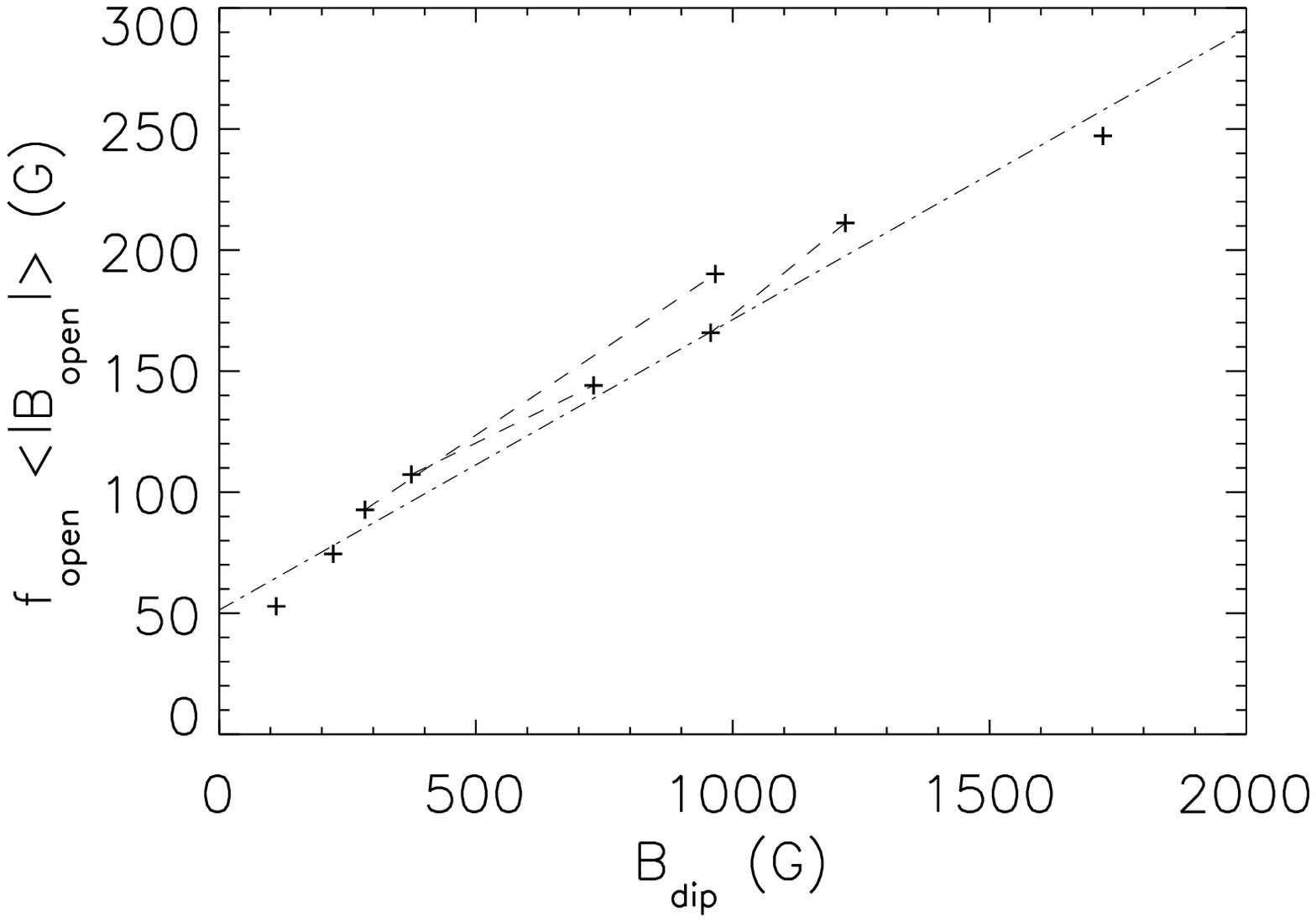}
\includegraphics[width=0.49\textwidth]{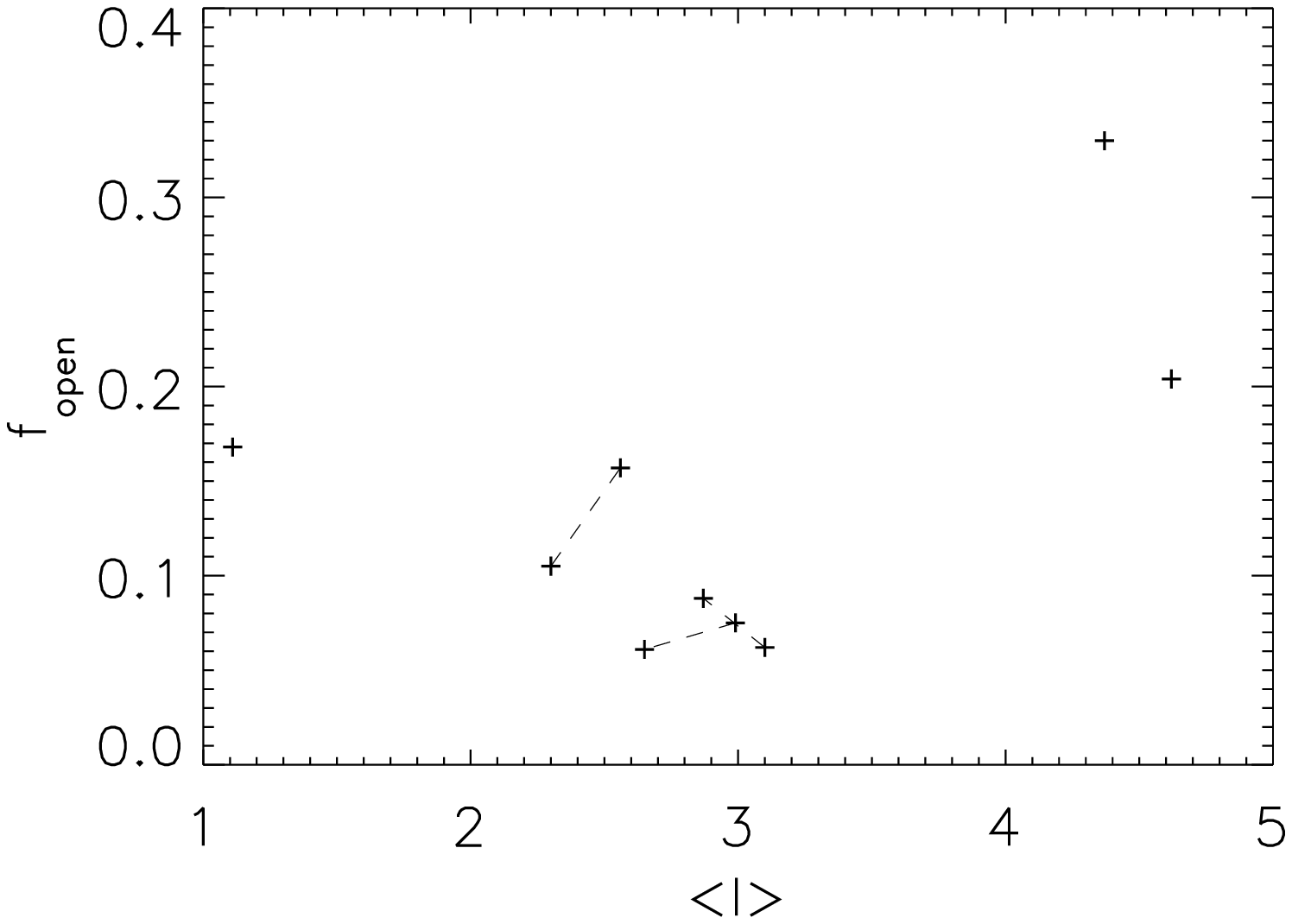}
\caption{
\emph{Upper panel}: correlation between the strength of the dipole component and $f_{open} <|B_{r,open}|>$, where $f_{open}$ is the fraction of the stellar surface covered in open flux, and $<|B_{r,open}|>$ is the average radial field strength in regions of open flux over the stellar surface. 
By definition, \mbox{$f_{open} <|B_{r,open}|> = \Phi_{open} / 4\pi R_\star^2$}, where $\Phi_{open}$ is the unsigned open magnetic flux.
The data is well fit by the relation \mbox{$f_{open} <|B_{r,open}|> = 0.12 B_{dip} + 51.3$ G}.
\emph{Lower panel}: correlation between the fraction of the stellar surface covered in regions of open flux, $f_{open}$, and the parameter $<l>$, which is related to the complexity of the magnetic field and the rate at which the field strength falls off with increasing distance from the star.
Data points representing the same star from two different epochs are connected by dashed lines. 
}
 \label{fig:openfluxresults}
\end{figure}

\subsection{Results}

For processes that involve the 3D magnetic field above the surface of the star, one of the most important parameters is the rate at which the field strength decreases with increasing distance from the star.
As we demonstrate in \mbox{Fig. \ref{fig:ratedropoff}}, this is related to how much magnetic energy is held within high $l$-value spherical harmonic components, which we measure using the parameter $<l>$ described in Section \ref{sect:surfacemaps}. 
When more magnetic energy is held in higher $l$-value modes, the magnetic field falls off with radius quickly close to the star. 
The rates at which all three magnetic fields decrease with distance become equal far from the stellar surface due to the field being dominated by the dipole component at these radii in all three cases.

The results of the coronal modelling for the stars in the sample are summarised in \mbox{Table \ref{tbl:cttscoronaresults}}.
We find that the closed coronae typically extend several stellar radii from the star. 
The coronal extent is determined by the distance from the star where the magnetic pressure and the thermal pressure balance each other. 
In this model, due to the assumption that the plasma pressure at the base of the corona is proportional to the magnetic pressure, the extent of the closed corona does not depend on the strength of the magnetic field.
Instead, the rate at which the field decreases in strength with increasing distance from the star is the dominant factor.
In \mbox{Fig. \ref{fig:extentcomplexity}}, we show that there is an anti-correlation between the closed coronal extent and the field complexity for the stars in the sample. 
Simpler fields are able to hold onto the hot coronal plasma much further from the stellar surface than more complex fields. 
\mbox{AA Tau}, which has the simplest field in the sample, has a closed corona that extends out to \mbox{4.5 R$_\star$}.
\mbox{V2247 Oph}, which has a much more complex field, has a closed corona that only extends out to 2 R$_\star$.

The other factors that determine the closed coronal extent are the stellar mass, the coronal temperature, and the rotation rate (\mbox{\citealt{1999A&A...346..883J}}; \mbox{\citealt{2006MNRAS.367..917J}}).
Decreasing the stellar mass leads to a larger surface gravity and a smaller pressure scale height; increasing the coronal temperature or the rotation rate leads to an increased pressure scale height\footnotemark.
Due to the slow rotation rates of the CTTSs considered here, centrifugal stripping does not have a significant effect on the sizes of the closed coronae of these stars.  
The lack of clear correlations between coronal temperature and closed coronal extent, and between stellar mass and closed coronal extent indicates that coronal temperature and stellar mass are not as important in determining the closed coronal extent as magnetic field complexity.

\footnotetext{
Increasing the stellar rotation rate also moves the corotation radius inwards. 
This can be significant for the closed coronal extents for rapidly rotating stars because inside of the corotation radius, the thermal pressure inside of a closed coronal loop decreases with increasing distance from the star, whereas outside of the corotation radius, the thermal pressure increases with increasing distance from the star (\mbox{\citealt{1999A&A...346..883J}}). 
This effect is not taken into account here because we assume in our coronal models that the source-surface is at the corotation radius, and therefore no closed coronal loops can extend past this radius.
}

For each of the stars, we calculate the emission measure weighted average coronal electron density for the modelled coronae and find values of $\bar{n}_e$ between $10^{8.7}$ cm$^{-3}$ and $10^{9.9}$ cm$^{-3}$.
As shown in \mbox{Fig. \ref{fig:extentcomplexity}}, the stars with more complex fields, and more compact coronae, have much higher average electron densities.
There is an order of magnitude variation between the minimum and the maximum electron densities in the sample.
This is similar to the result of \mbox{\citet{2006MNRAS.367..917J}} who found that stars with simple field structures tend to have average coronal electron densities that are an order of magnitude lower than stars with more complex fields. 
The absence of magnetic flux on small-scales in the ZDI magnetic maps can lead to an underestimation of the modelled electron densities (\mbox{\citealt{2010MNRAS.404..101J}}).

For each of the stars, our calculations of the open unsigned magnetic flux is based on both the PFSS model and the plasma model.
We assume a field line is open if it extends from the stellar surface to the source-surface, or if the plasma beta is greater than unity at any point along its length.
The open unsigned magnetic flux can be written as a function of two parameters: these are the average radial field strength at the stellar surface in regions of open field, $<|B_{r,open}|>$, and the area of the stellar surface covered in open field, $4 \pi R_\star^2 f_{open}$, where $f_{open}$ is the open flux filling factor. 
The open flux filling factor can have an effect on the properties of stellar winds launched from coronal holes.
By analogy with the solar wind, larger values of $f_{open}$ lead to smaller coronal expansion factors (i.e. smaller rates of expansion of magnetic flux tubes with increasing distance from the star) which can lead to stellar winds that are slower and more dense (\mbox{\citealt{1990ApJ...355..726W}}; \mbox{\citealt{2000JGR...10510465A}}; \mbox{\citealt{2008JGRA..11308112M}}). 
The open magnetic flux is given by

\begin{equation}
\Phi_{open} = 4 \pi R_\star^2 f_{open} <|B_{r,open}|>.
\end{equation}

\noindent The values of $f_{open}$ and $<|B_{r,open}|>$ are given in \mbox{Table \ref{tbl:cttscoronaresults}}. 
In \mbox{Fig. \ref{fig:openfluxresults}} we show that the value of $f_{open} <|B_{r,open}|>$  is determined by the strength of the dipole component of the field. 
This means that for each star, the open unsigned magnetic flux is determined by the strength of the dipole component and the surface area of the star. 
The following relation is derived from the linear fit to the results presented in \mbox{Fig. \ref{fig:openfluxresults}}:

\begin{equation}
\Phi_{open} R_\star^{-2} = 1.5 B_{dip} + 644.0 \textrm{ G}.
\end{equation}

\begin{figure*}
\subfigure[AA Tau 2009]{\includegraphics[width=0.32\textwidth]{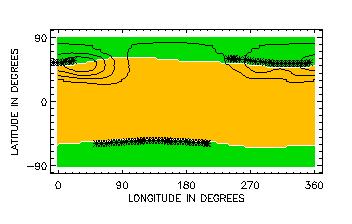}}
\subfigure[BP Tau Feb 2006]{\includegraphics[width=0.32\textwidth]{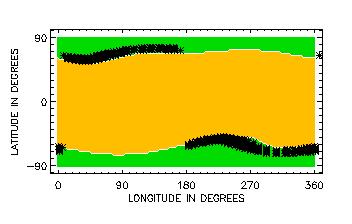}}
\subfigure[V2129 Oph 2009]{\includegraphics[width=0.32\textwidth]{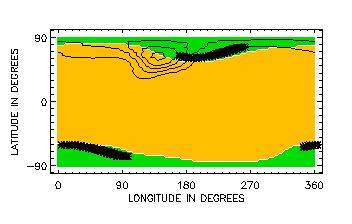}}
\subfigure[BP Tau Dec 2006]{\includegraphics[width=0.32\textwidth]{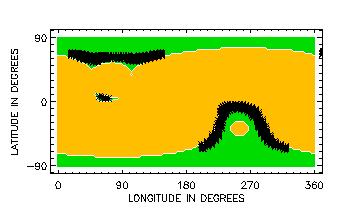}}
\subfigure[TW Hya 2010]{\includegraphics[width=0.32\textwidth]{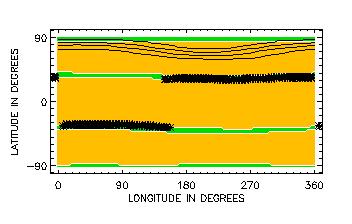}}
\subfigure[TW Hya 2008]{\includegraphics[width=0.32\textwidth]{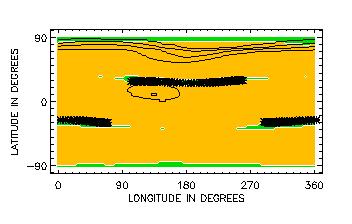}}
\subfigure[V2129 Oph 2005]{\includegraphics[width=0.32\textwidth]{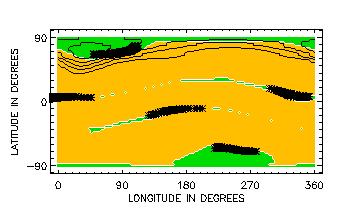}}
\subfigure[CR Cha]{\includegraphics[width=0.32\textwidth]{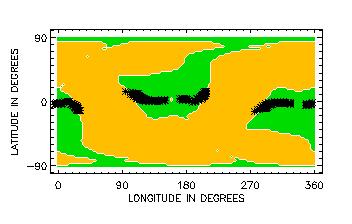}}
\subfigure[V2247 Oph]{\includegraphics[width=0.32\textwidth]{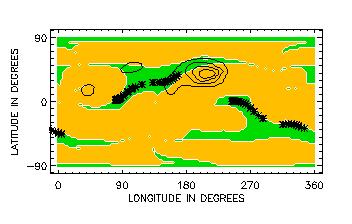}}
\caption{
Surface maps showing the predicted locations of open magnetic field (green), closed magnetic field (yellow) and accretion footpoints (black) for most of the stars in the sample. 
The black contour lines show the locations of excess Ca II IRT emission inferred from Doppler Imaging; this is considered a good observational constraint on the locations of accretion footpoints.
These maps are arranged from left to right and from top to bottom by decreasing strength of the dipole components of the field.
}
 \label{fig:CTTSaccdisp}
\end{figure*}

The locations of open and closed field structures for each of the stars in the sample are shown in \mbox{Fig. \ref{fig:CTTSaccdisp}}.
The distributions of these structures are a strong function of field complexity, with regions of open flux covering the magnetic poles on all stars.
Simple axisymmetric dipole fields such as the one seen on \mbox{AA Tau} have regions of open flux predominantly at high latitudes, with all low-latitude field structures being closed, and regions of open flux covering 10-20\% of the stellar surface.
More complex fields, such as that seen on \mbox{V2247 Oph}, show open flux distributed over all latitudes, covering $\approx$ 20\% of the stellar surface. 
Fields that are dominated by large-scale octupole components, such as those seen on \mbox{TW Hya} and \mbox{V2129 Oph}, tend to have regions of open flux at mid-latitudes extending in bands around the star, with only 5-10\% of the stellar surface covered in regions of open field.
The range of values of the open flux filling factors found here are in good agreement with those found for the Sun by applying the PFSS model to observed surface magnetic field maps (\mbox{\citealt{1990ApJ...355..726W}}).

\begin{figure*}
\includegraphics[width=0.49\textwidth]{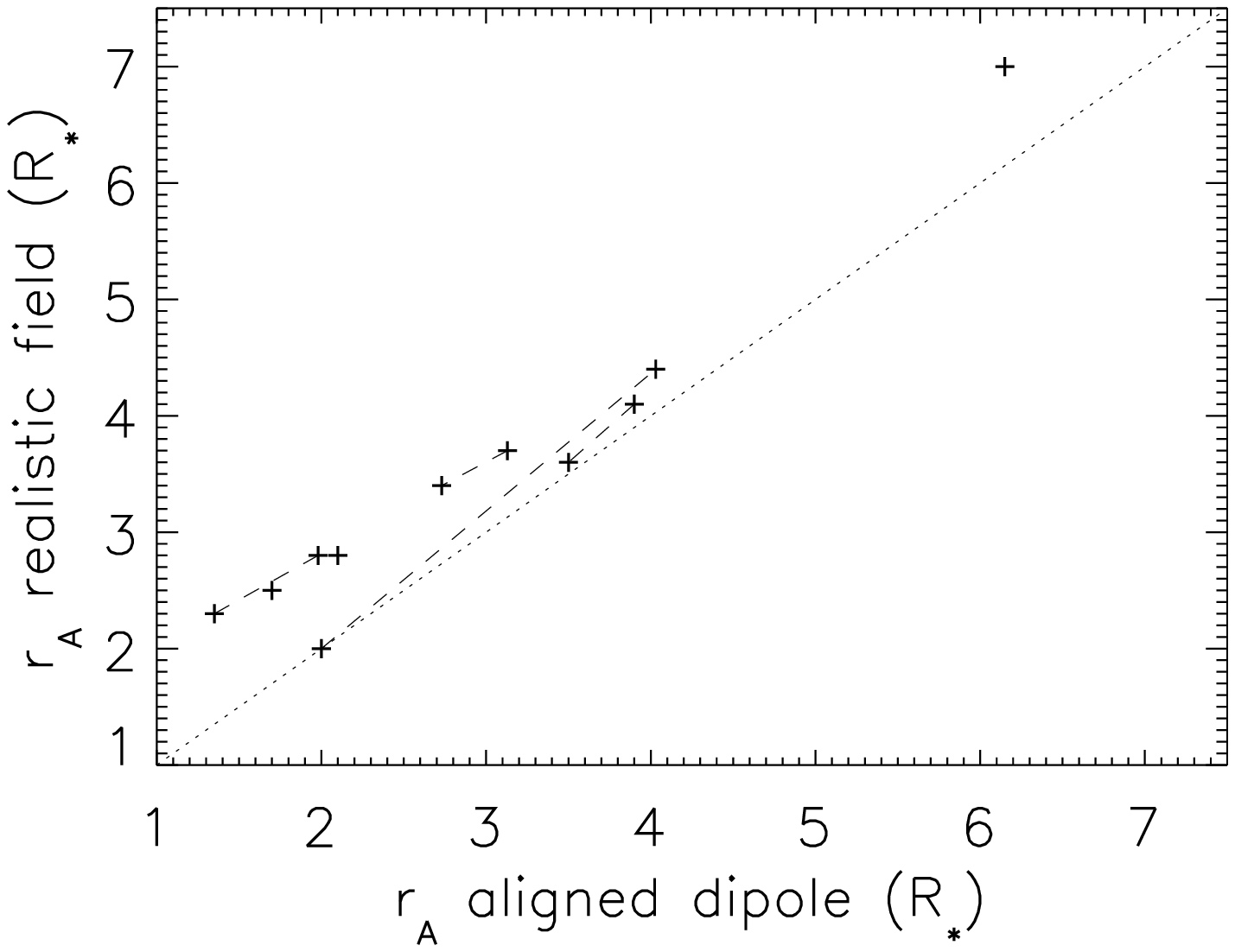}
\includegraphics[width=0.49\textwidth]{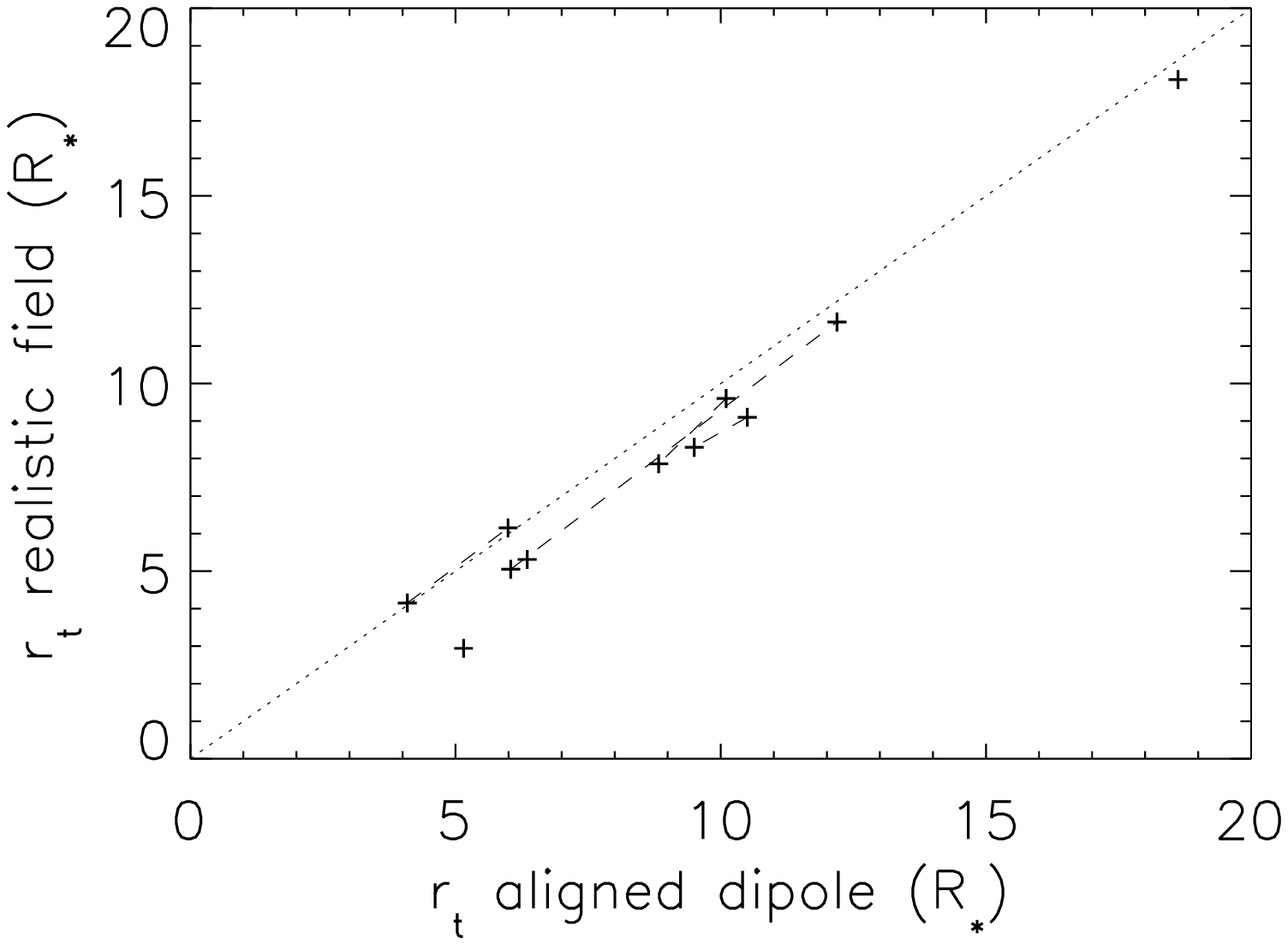}
\includegraphics[width=0.49\textwidth]{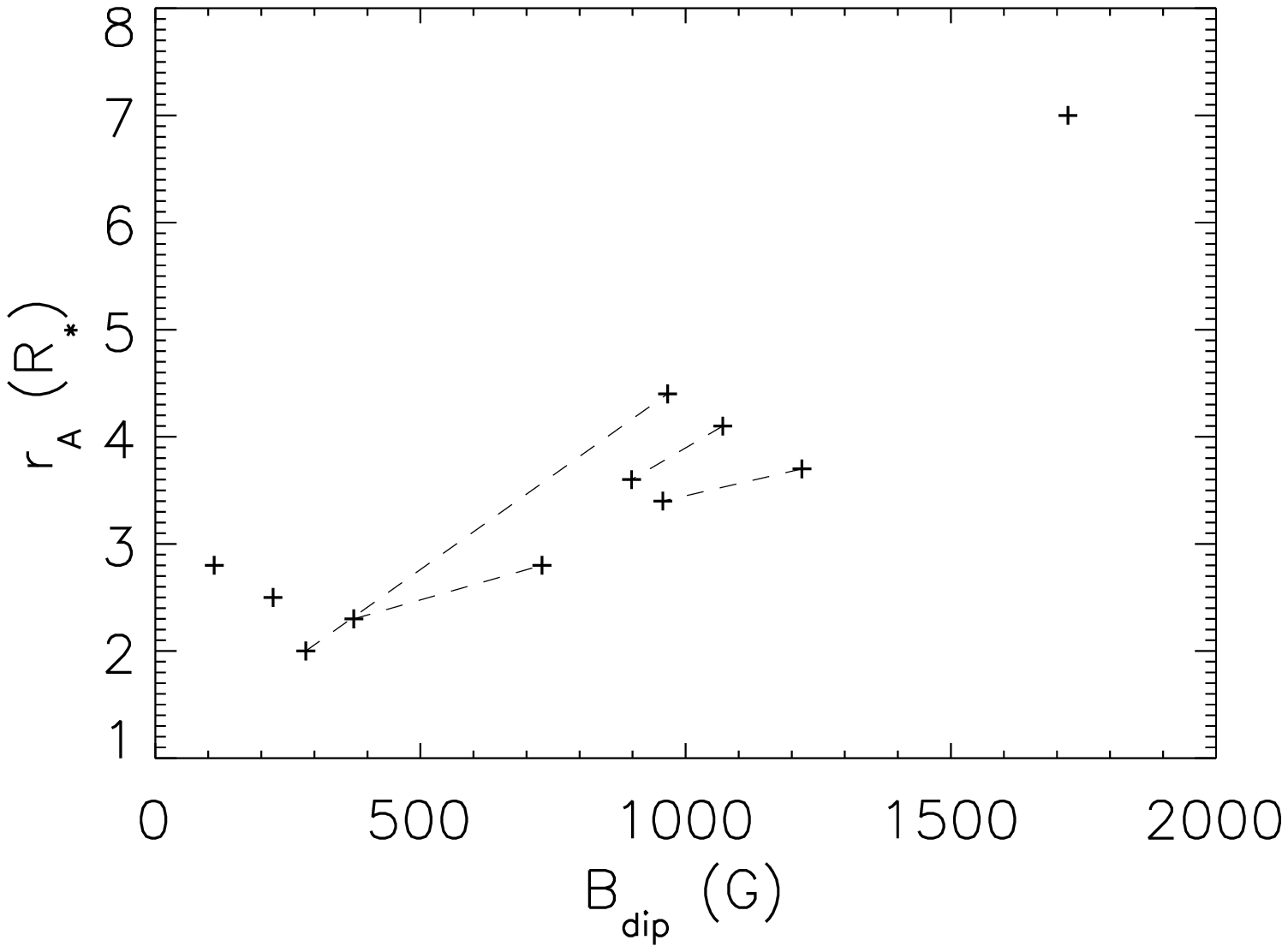}
\includegraphics[width=0.49\textwidth]{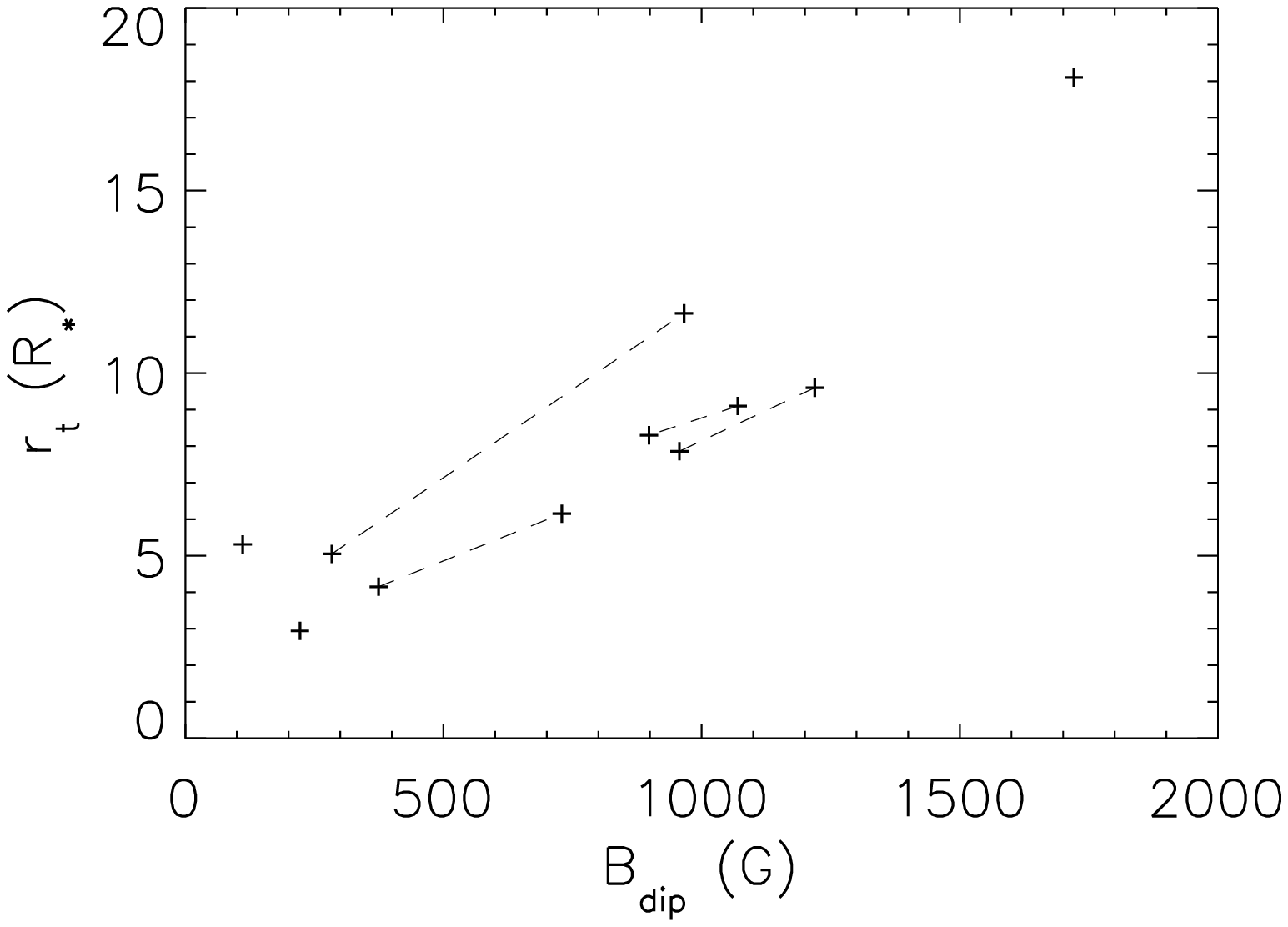}
\caption{
\emph{Upper-left panel}: Alfv\'{e}n surface radii calculated from realistic field structures using \mbox{Eqn. \ref{eqn:alfven2}} against the values calculated by assuming that the fields are axisymmetric dipoles using \mbox{Eqn. \ref{eqn:discalfvenradius}}.
\emph{Lower-left panel}: Alfv\'{e}n surface radii calculated from realistic field structures against the strengths of the dipole components of the fields. 
\emph{Upper-right panel}: torque balance radii calculated from realistic field structures using \mbox{Eqn. \ref{eqn:torquebalance}} against the values calculated by assuming that the fields are axisymmetric dipoles using \mbox{Eqn. \ref{eqn:torquebalance2}}.
\emph{Lower-right panel}:  torque balance radii calculated from realistic field structures against the strengths of the dipole components of the fields. 
Data points representing the same star from two different epochs are connected by dashed lines.
\mbox{CV Cha}, \mbox{V4046 Sgr A}, and \mbox{V4046 Sgr B} are not included.
}
\label{fig:ralfvenstuff}
\end{figure*}

\section{Disc truncation and accretion} \label{sect:acc}

\subsection{Model}

In this section, we use a simple model to investigate disc truncation and magnetospheric accretion geometries.
We assume that the disc is truncated by the stellar magnetic field and accretion occurs along all field lines that connect to the disc between the disc truncation radius and the outer radius at which that magnetic field can begin to deflect material out of the disc plane and into an accretion column.
We therefore need to estimate the disc truncation radius and the outer radius at which the magnetic field is able to disrupt the disc for each star in the sample. 

We assume that the disc truncation radius is at the Alfv\'{e}n radius where the magnetic energy density is balanced by the kinetic energy density from the disc: this is where $\rho(r) v(r)^2 /2 = B(r)^2 / 8\pi$, where $\rho$ is the mass density of material in the disc, $v$ is the velocity of material in the disc (approximately equal to the Keplerian orbital velocity in this case), and $B$ is the stellar magnetic field strength.
Making this assumption, \mbox{\citet{1977ApJ...215..897E}} derived an expression for the case of isotropic spherical accretion of material falling at free-fall speeds onto a star.
Conservation of mass and the isotropic density distribution mean that the mass accretion rate is given by $\dot{M}_a = 4 \pi r^2 \rho(r) v(r)$, and the free-fall assumption means that the infall velocity of accreting material is given by $v(r) = \left( G M_\star / r \right)^{1/2}$.
This implies that the Alfv\'{e}n surface is where

\begin{equation} \label{eqn:alfven2}
B^2 = C (G M_\star)^{\frac{1}{2}}  \dot{M}_{a}  r_A^{-\frac{5}{2}}
\end{equation}

\noindent where $r_A$ is the Alfv\'{e}n radius, and $C$ is a factor (not present in the \mbox{\citet{1977ApJ...215..897E}} formulation) that accounts for the difference between magnetospheric accretion from a disc and spherical accretion.
\mbox{\citet{1977ApJ...215..897E}} assumed that the star's magnetic field is a simple dipole, such that $B = \mu_1 r^{-3}$, where $\mu_1=B_\star R_\star^3$ and  $B_\star$ is the stellar surface magnetic field strength (in the case of accretion from a disc with a axisymmetric dipole field, this can be taken as the field strength in the equatorial plane). 
Substituting this into \mbox{Eqn. \ref{eqn:alfven2}} gives the familiar expression
 
\begin{equation} \label{eqn:discalfvenradius}
r_A = 2^{-\frac{1}{7}} k \left[   \frac{ B_\star^2 R_\star^6}{(G M_\star)^{\frac{1}{2}} \dot{M}_a}   \right]^{2/7}
\end{equation}

\noindent where $k$ is another factor that accounts for the difference between magnetospheric accretion from a disc and spherical accretion and is given be $k = (2^{-1/2} C)^{-2/7}$.
\mbox{\citet{2005ApJ...634.1214L}} used MHD accretion simulations to estimate \mbox{$k \sim 0.5$} for magnetospheric accretion from a disc. 
We assume $C \sim 16$ in \mbox{Eqn. \ref{eqn:alfven2}}, because this is equivalent to \mbox{$k \sim 0.5$} in \mbox{Eqn. \ref{eqn:discalfvenradius}}.
For each of the stars, we solve \mbox{Eqn. \ref{eqn:alfven2}} using field extrapolations without a source-surface and by taking $B^2$ at each radius as the longitudinal average value of $B^2$ in the equatorial plane.

An outer estimate for the radius at which the magnetic field can affect accretion is the radius at which the magnetic torque on the disc is balanced by the internal viscous torque (e.g. \mbox{\citealt{2008A&A...478..155B}}); we call this the torque balance radius $r_t$.
Due to the difference in rotation rate between the disc and the stellar surface (except for at the corotation radius), the magnetic field lines connecting the star and the disc become twisted.
As in previous studies (e.g. \mbox{\citealt{1995MNRAS.273..639C}}; \mbox{\citealt{1996ApJ...465L.111W}}; \mbox{\citealt{2008MNRAS.389.1839G}}) we assume that this twist is $45^\circ$, and so $|B_z| = |B_\phi^+|$, where $B_\phi^+$ is the longitudinal component of the magnetic field that is generated by the twisting of magnetic field lines by the inner edge of the circumstellar disc.
In the equatorial plane $B_z=B_\theta$, so the torque balance radius can be found by solving

\begin{equation} \label{eqn:torquebalance}
B_\theta^2 r_t^2 = \frac{1}{2} \dot{M}_{a} \left( \frac{G M_\star}{r_t} \right)^{\frac{1}{2}}
\end{equation}

\noindent where we find $r_t$ using field extrapolations without a source-surface and taking $B_\theta^2$ at each radius as the value of $B_\theta^2$ averaged over all longitudes in the equatorial plane.
As in \mbox{\citet{1995MNRAS.273..639C}}, if we assume that the field is an axisymmetric dipole, such that $B_\theta = B_\star R_\star^3 r^{-3}$, the torque balance radius is given by

\begin{equation} \label{eqn:torquebalance2}
r_t = 2^{\frac{2}{7}} \left[   \frac{B_\star^2 R_\star^6}{(G M_\star)^{\frac{1}{2}} \dot{M}_a}   \right]^{\frac{2}{7}}.
\end{equation}

Since accretion cannot occur onto the star from regions of the disc outside of the corotation radius, in cases where we calculate $r_t$ to be outside the corotation radius, we take the outer radius at which accretion can occur to be the corotation radius.

In several cases, the torque balance radii and the Alfv\'{e}n radii will be overestimated because the quantity $\dot{M}_a$ used in the two above equations is a measure of the rate at which mass moves inwards through the disc (introduced into the equations because the viscosity in the disc is not well understood), and not a measure of the rate at which material accretes onto the central star. 
Since some disc material will be lost in the form of disc winds, and not accrete onto the central star, the measured mass accretion rates underestimate the rate at which mass moves inwards through the disc.

For comparison, we also calculate the disc truncation radii using the alternative formulation of \mbox{\citet{2008A&A...478..155B}}, given by 

\begin{equation} \label{eqn:bessolaz}
r_{trunc} = 2  m_s^{\frac{2}{7}}    \left( \frac{B_\star}{140}  \right)^{\frac{4}{7}}      \left( \frac{\dot{M}_a}{10^{-8}}   \right)^{-\frac{2}{7}} \left(\frac{M_\star}{0.8}   \right)^{-\frac{1}{7}}  \left( \frac{R_\star}{2}   \right)^{\frac{5}{7}} 
\end{equation}

\noindent where $m_s \sim 1$, and the quantities $B_\star$, $\dot{M}_a$, $M_\star$, and $R_\star$ are in units of G, M$_\odot$ yr$^{-1}$, M$_\odot$, and R$_\odot$ respectively. 
The results from these calculations can be seen and compared with the $r_A$ and $r_t$ estimates in \mbox{Table \ref{tbl:CTTSrinroutresults}} and \mbox{Fig. \ref{fig:radiiresults}}.

In order to estimate $r_A$, $r_t$ and $r_{trunc}$ for each star, we adopt the same mass accretion rates as used in the papers where the magnetic maps were presented, as listed in \mbox{Table \ref{tbl:stellarparams}}.
These values were calculate from the same observations that were used to derive the magnetic maps and are therefore the most appropriate for these calculations.

When the inner and outer radii from which accretion can occur are estimated, we use PFSS extrapolations with the source-surface radius set to the corotation radius, and trace field lines from every point in the disc from which accretion can occur and assume that all such field lines support accretion.
We assume that accretion only occurs along the field line in the direction at which the component of effective gravity along the field line in a coordinate system corotating with the star points inwards towards the stellar surface. 

\mbox{\citet{2010MNRAS.404..101J}} showed that the possible absence of magnetic flux on small-scales in the ZDI magnetic maps is unlikely to have a significant effect on accretion models.
This is because accretion is affected primarily by the large-scale structure of the stellar magnetic field, with the small-scale structures having little significance.

\begin{table*}
\centering
\begin{tabular}{ccccccccc}
\hline 
Star & $r_{co}$ ($R_\star$) & $r_{A}$ ($R_\star$) & $r_{A}$ ($R_\star$) & $r_{t}$ ($R_\star$) & $r_{t}$ ($R_\star$) & $r_{trunc}$ ($R_\star$)\\
 & & Dipole & Realistic & Dipole & Realistic & Bessolaz \\
\hline

AA Tau 2009 &			7.6 & 		6.9 & 	7.0 & 	18.6 &	18.1 &	12.7\\

BP Tau Feb 2006 & 		6.0 &			3.8 & 	3.7 & 	10.1 &	9.6 &		6.9\\

GQ Lup 2009 &			10.4 &		3.9 &		4.1 &		10.5 &	9.1 &		7.1\\

V2129 Oph 2009 & 		8.1 &			4.5 & 	4.4 & 	12.2 &	11.6 &	8.3\\

BP Tau Dec 2006 & 		6.0 &			3.3 & 	3.4 & 	8.8 &		7.9 &		6.0\\

GQ Lup 2011 &			10.4 &		3.5 &		3.6 &		9.5 &		8.3 &		6.4\\

TW Hya 2010 & 		8.3 &			2.2 & 	2.8 & 	6.0 &		6.1 &		4.1\\

TW Hya 2008 & 		8.3 &			1.5 & 	2.3 & 	4.1 &		4.2 &		2.8\\

V2129 Oph 2005 & 		8.1 &			2.2 & 	2.0 & 	6.0 &		5.1 &		4.1\\

CR Cha & 			3.6 &			1.9 & 	2.5 & 	5.2 &		2.9 &		3.5\\

V2247 Oph & 			3.4 &			2.4 & 	2.8 & 	6.3 &		5.3 &		4.3\\

\hline
\end{tabular}
\caption{
Results of the disc truncation radius calculations. 
The maps are arranged from top to bottom by descending strength of the dipole component of the field.
From left to right, the columns correspond to the Alfv\'{e}n surface radius calculated assuming the field is an axisymmetric dipole (\mbox{Eqn. \ref{eqn:discalfvenradius}}), the Alfv\'{e}n radius calculated using potential field extrapolations of the ZDI magnetic maps (\mbox{Eqn. \ref{eqn:alfven2}}), the torque balance radius calculated assuming the field is an axisymmetric dipole (\mbox{Eqn. \ref{eqn:torquebalance2}}), the torque balance radius calculated using potential field extrapolations of the ZDI magnetic maps (\mbox{Eqn. \ref{eqn:torquebalance}}), and the alternative formula (\mbox{Eqn. \ref{eqn:bessolaz}}) of \mbox{\citet{2008A&A...478..155B}}.
}
\label{tbl:CTTSrinroutresults}
\end{table*}

\begin{figure}
\includegraphics[width=0.49\textwidth]{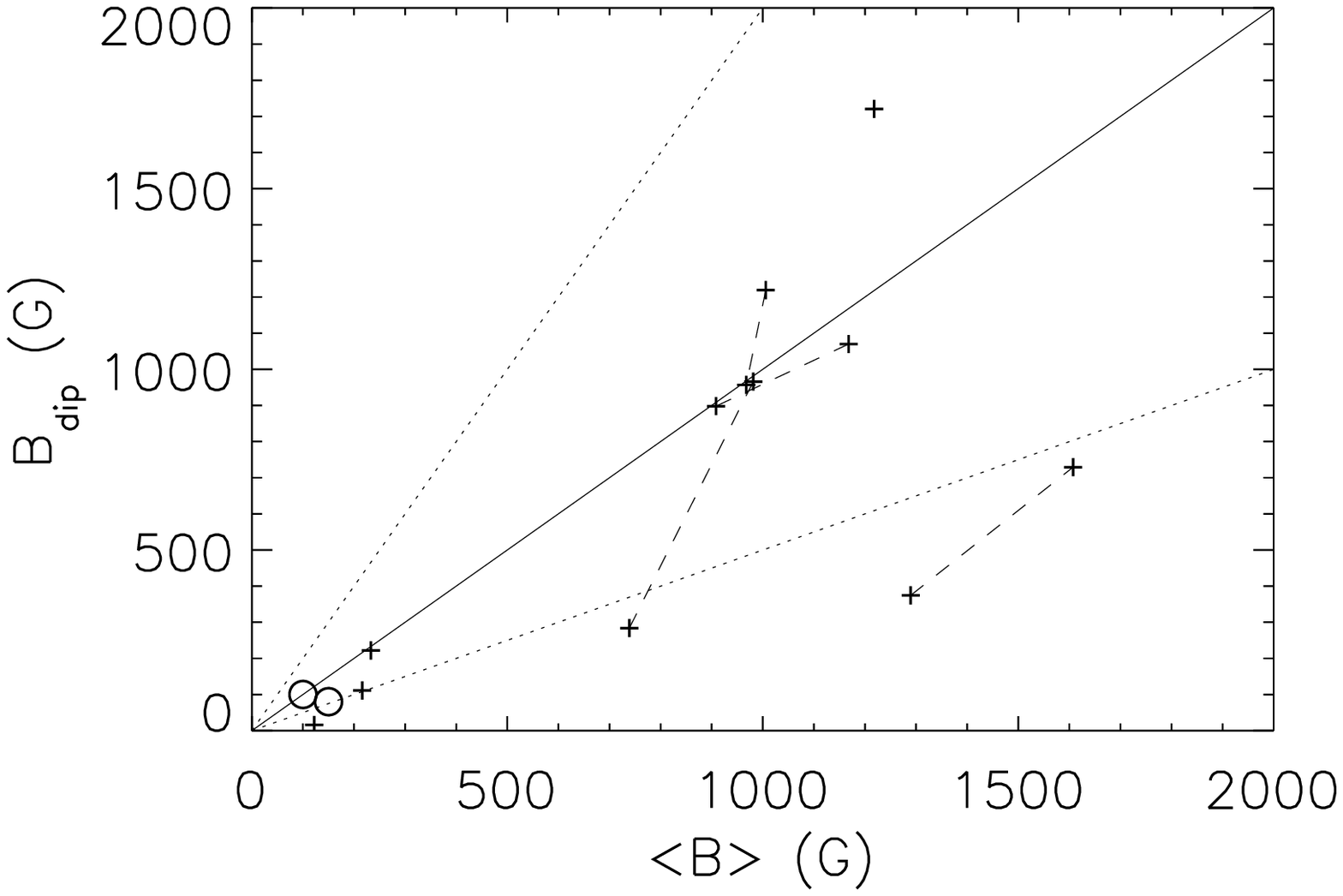}
\caption{
Dipole field strength against surface averaged field strength for the sample of CTTS ZDI maps.
The solid line shows the line of equality, and the dotted lines show the lines of \mbox{$B_{dip} = 2 <B>$} and \mbox{$B_{dip} = <B>/2$}.
The two stars in the binary system \mbox{V4046 Sgr} are shown as open circles.  
Data points representing the same star from two different epochs are connected by dashed lines.
The surface averaged field strengths should be considered as lower limits because they do not take into account missing magnetic flux on small-scales.
This figure shows that the surface averaged magnetic field strengths are not reliable indicators of the dipole field strength. 
}
\label{fig:BdipvsBave}
\end{figure}

\subsection{Results}

In \mbox{Table \ref{tbl:CTTSrinroutresults}}, we give the values of the Alfv\'{e}n surface radii and the torque balance radii for each of the stars. 
These have both been calculated twice: first by assuming that the field is an axisymmetric dipole with the polar field strength of the dipole taken to be that of the dipole component derived from the ZDI magnetic maps, and then using the potential-field extrapolations of the ZDI magnetic maps.
In \mbox{Fig. \ref{fig:ralfvenstuff}}, we show the relation between the results calculated using the two methods (i.e. using the assumption of an axisymmetric dipole field, and using the potential field extrapolations from the ZDI maps) for both the Alfv\'{e}n surface radii and the torque balance radii.
In general, the values are similar, indicating that it is reasonable to approximate the field as an axisymmetric dipole when calculating the disc truncation radius using the analytic formulations described above, though it is necessary to consider higher order spherical harmonic components when describing the flow of matter onto the stellar surface (\mbox{\citealt{2011AN....332.1027G}}; \mbox{\citealt{2012ApJ...744...55A}}).
However, the disc truncation radius can only be estimated if the strength of the dipole component of the field is accurately known.
In many cases, what is actually known is the surface averaged field strength that takes into account every component of the field, and not just the dipole.
In \mbox{Fig. \ref{fig:BdipvsBave}}, we show that there is only a weak correlation between the surface averaged magnetic field strength derived from the ZDI map and the strength of the dipole component of the field. 
This figure does not take into account the fact that most of the magnetic flux on small-scales that could be measured by Zeeman broadening techniques is not captured in the ZDI maps.
It is not possible to accurately estimate the strength of the dipole component of a field, and therefore the disc truncation radius, if the only quantity known is the surface averaged field strength.

The differences between the Alfv\'{e}n surface radii calculated using the two methods is a result of two factors: the tilt of the dipole component of the field relative to the rotation axis and the presence of the octupole component.
The tilt in the dipole component increases the field strength in the equatorial plane relative to what it would be for an axisymmetric dipole.
This pushes the inner edge of the disc away from the star.
This is signficant for the \mbox{CR Cha} field and the \mbox{TW Hya 2008} field because they have significantly tilted dipole components.
The effect of the octupole component of the field depends on whether it is aligned or anti-aligned with the dipole component (i.e. whether the positive magnetic pole of the octupole component is in the same hemisphere as the positive magnetic pole or the negative magnetic pole of the dipole component).
If the octupole component is aligned with the dipole, the two components will add up destructively in the equatorial plane, leading to a weaker field and a smaller Alfv\'{e}n surface radius.
If the octupole component is anti-aligned with the dipole, the two components add up constructively in the equatorial plane, leading to a stronger field and a larger Alfv\'{e}n surface radius.
The effects of parallel and anti-parallel dipole and octupole components are discussed in detail in \mbox{\citet{2011AN....332.1027G}} and \mbox{\citet{2012ApJ...744...55A}}.

In general, the discs are truncated several stellar radii from their stars, with \mbox{AA Tau} having the largest truncation radius.
This is unsurprising given \mbox{AA Tau}'s strong dipolar magnetic field. 
The truncation radius for \mbox{AA Tau} is close to the corotation radius, which is consistent with the observations of \mbox{\citet{1999A&A...349..619B}} and \mbox{\citet{2003A&A...409..169B}} that showed that the star is being eclipsed periodically by a warped inner edge of its disc, with the warp located at approximately the corotation radius.

We show in \mbox{Fig. \ref{fig:ralfvenstuff}} that stars with weaker dipole components tend to have smaller Alfv\'{e}n surface radii; however the stellar mass, stellar radius, and mass accretion rate are also very important in determining the Alfv\'{e}n surface radius. 
For \mbox{CV Cha}, which has a weak dipole component and a large measured mass accretion rate, the magnetic field is not able to truncate the disc outside of the star, and so we do not consider it here.

The torque balance radii are generally significantly larger than the Alfv\'{e}n surface radii.
The exception to this is \mbox{CR Cha}, where the large tilt in the dipole component means that the method used for calculating the torque balance radii is not valid because the method assumes that the magnetic field lines threading the disc are vertical prior to being disturbed by the orbital motion of the disc material, which is not the case for a significantly tilted dipole.
For \mbox{AA Tau} especially, due to its strong axisymmetric dipole component, the torque balance radius is very large, indicating that \mbox{AA Tau}'s magnetic field may be able to disrupt the disc well outside the corotation radius.
For \mbox{V2129 Oph}, the 2005 field has a torque balance radius inside the corotation radius, but due to the large increase in the strength of the dipole component, the 2009 field has a torque balance radius that has moved outside of the corotation radius. 

In \mbox{Fig. \ref{fig:CTTSaccdisp}}, we show the locations of the predicted accretion footpoints for each of the stars in the sample. 
These locations are highly dependent on magnetic field structure.

\begin{figure*}
\includegraphics[width=0.4\textwidth]{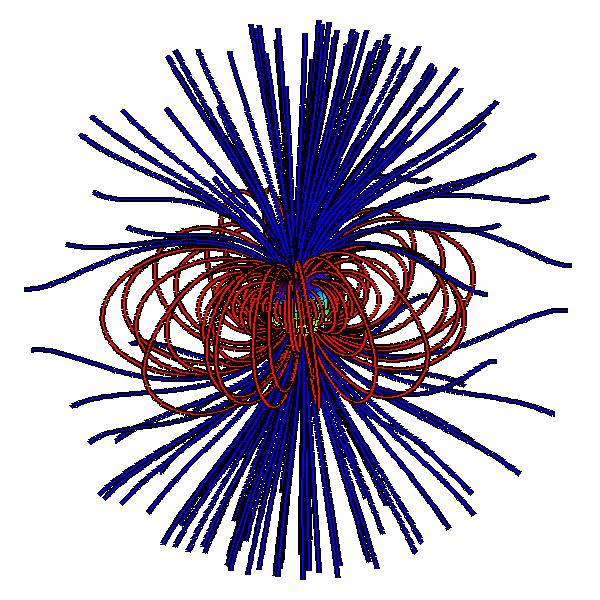}
\includegraphics[width=0.4\textwidth]{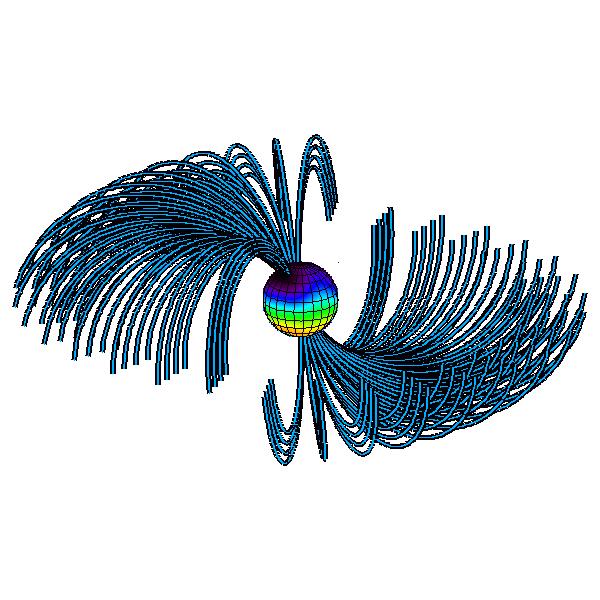}
\includegraphics[width=0.4\textwidth]{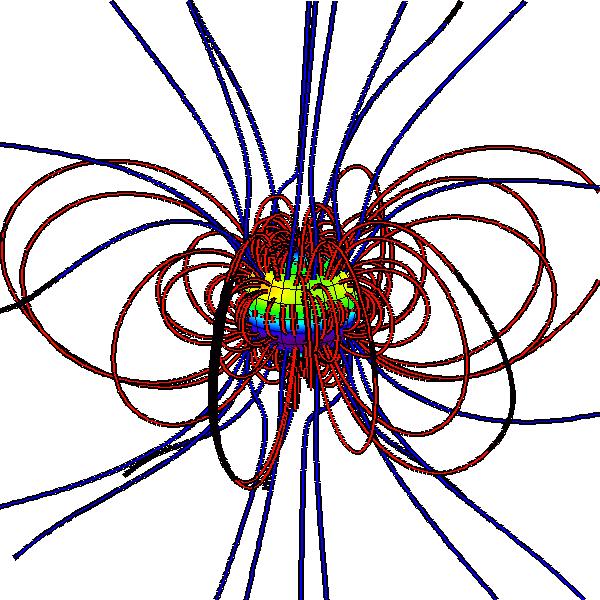}
\includegraphics[width=0.4\textwidth]{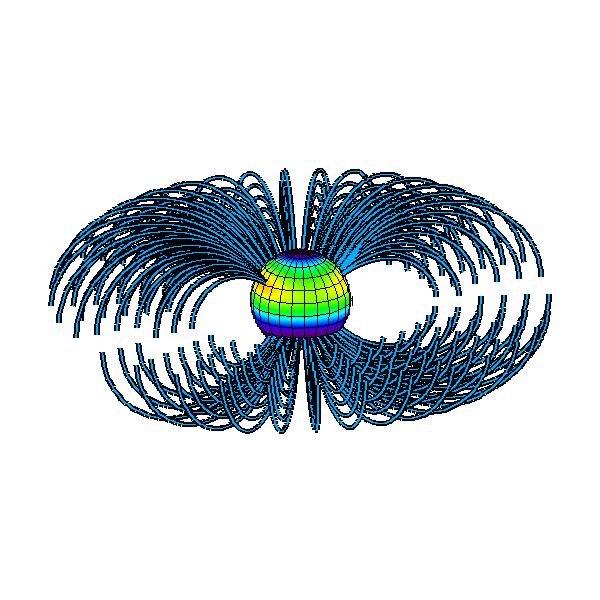}
\includegraphics[width=0.4\textwidth]{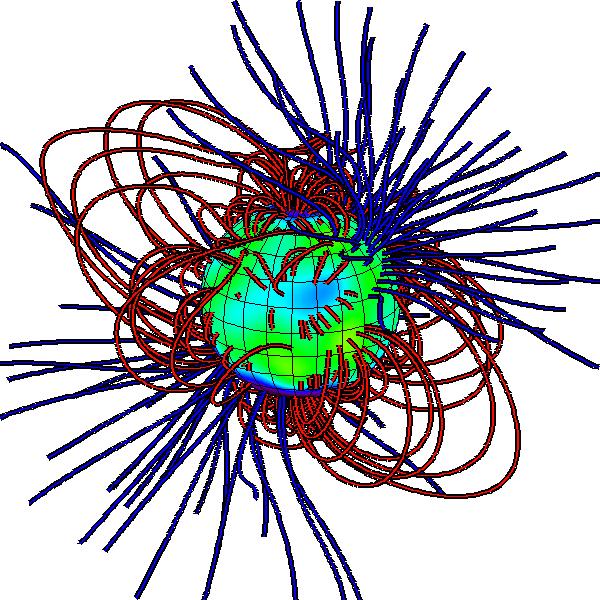}
\includegraphics[width=0.4\textwidth]{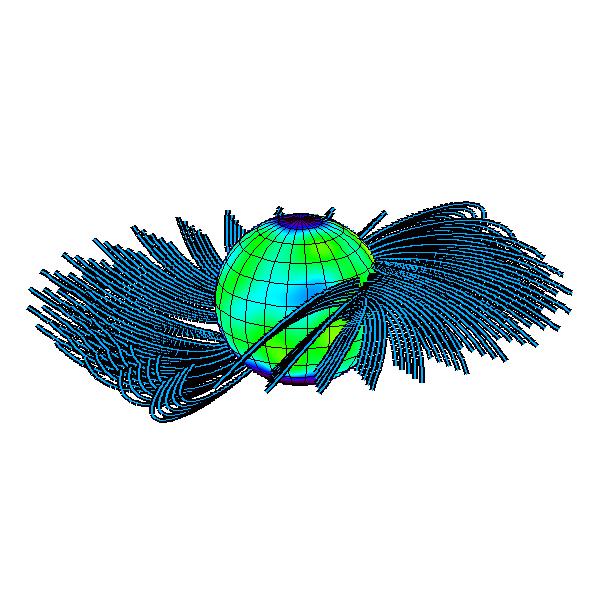}
\caption{Potential field extrapolations for \mbox{AA Tau} (\emph{upper panels}), the \mbox{TW Hya 2010} field (\emph{middle panels}), and \mbox{V2247 Oph} (\mbox{\emph{lower panels}}) showing closed and open field lines (\emph{left-hand panels}) and accreting field lines (\emph{right-hand panels}).
Open field lines are deep blue, closed field lines are red, and field lines that are able to support accretion are light blue.
These three stars have been chosen to represent the three kinds of accretion geometries that are typical in the sample of CTTSs.
The dominantly dipolar field of \mbox{AA Tau} leads to accretion footpoints at high latitudes, while the material accreting onto \mbox{TW Hya} is deflected towards mid-latitudes due to its strong octupole field component that is approximately anti-parallel to its weak dipole component. 
The complex field of \mbox{V2247 Oph} leads to a complex pattern of mid-to-low latitude accretion.
}
\label{fig:acclines}
\end{figure*}

For the most dipolar fields, such as the \mbox{AA Tau} field, the two \mbox{BP Tau} fields, and the \mbox{V2129 Oph 2009} field, the accretion footpoints are all at high latitudes. 
These are also the fields with the lowest values of $<l>$.
The simple fields also have the strongest dipole components, and therefore have the largest disc truncation radii.
As these fields have strong dipole components, and the discs are truncated far from the stellar surfaces, the fields are very dipolar at the inner disc edges, and accretion happens along large-scale field lines that connect with the stellar surface at high latitudes.
The field structure and accreting field lines for \mbox{AA Tau} are shown in \mbox{Fig. \ref{fig:acclines}}.

The two \mbox{TW Hya} fields are dominated by several kG octupole components and relatively weak dipole components, with the dipoles and octupoles close to an anti-parallel configuration (\mbox{\citealt{2011MNRAS.417..472D}}; \mbox{\citealt{2011AN....332.1027G}}). 
The fields lines that connect the stars with the inner edges of the discs are predominantly dipolar. 
However, since both these fields have octupole components that are significantly stronger than the dipole components near the stellar surface, the large-scale field lines carrying the accretion flow onto the star are deflected towards mid-latitudes. 
This is the opposite behaviour to what is found for stars with dipole and octupole moments that are close to parallel, where the accretion flow is instead deflected towards higher latitude than would be expected for a pure dipole field (see \mbox{\citealt{2011AN....332.1027G}} where the accretion geometries for dipole-octupole magnetic fields are discussed in detail).
The field structure and accreting field lines for the \mbox{TW Hya 2010} field are shown in \mbox{Fig. \ref{fig:acclines}}.

For the most complex fields, accretion impacts the surface at low-latitudes and is distributed over the stellar surface in a much more complex pattern.
This is especially true for \mbox{CR Cha} and \mbox{V2247 Oph}, both of which have highly complex fields.
As \mbox{V2247 Oph} has a weak dipole component, its disc is truncated close to the stellar surface despite its low mass accretion rate.
Thus, the field is highly non-dipolar at the inner edge of the disc and accretion occurs along field lines that connect with the stellar surface at mid-to-low latitudes and in a complex pattern (c.f. \mbox{\citealt{2011AN....332.1027G}}).
The field structure and accreting field lines for \mbox{V2247 Oph} are shown in \mbox{Fig. \ref{fig:acclines}}.

In many cases, the locations of the accretion footpoints estimated using our simplified model are different from the locations of accretion footpoints estimated using Doppler Imaging maps Ca II infrared triplet excess emission.
The locations of this excess emission are shown for six of the stars as black contour lines in \mbox{Fig. \ref{fig:CTTSaccdisp}}.
For the cases of \mbox{AA Tau} and the \mbox{V2129 Oph 2005} epoch, the predicted accretion footpoint locations are in good agreement with the observational constraints.
For the cases of \mbox{V2247 Oph} and the \mbox{V2129 Oph 2009} epoch, the predictions and observations are similar, but the strongest observed regions of accretion are shifted in longitude by a few tens of degrees from the predicted accretion footpoint locations. 
For the two \mbox{TW Hya} epochs, there is a poor match between predictions and observations, with the models predicting that accretion streams impact the star near the equator and observations suggesting that the accretion streams impact the star near the poles. 
A similar disagreement between observations and models has previously been noted for the case of \mbox{BP Tau} by \mbox{\citet{2008MNRAS.386.1234D}}.
\mbox{\citet{2008MNRAS.386.1234D}} showed that this problem could be solved by introducing a model where the inner edge of the disc is warped, such that instead of lying entirely in the stellar equatorial plane, the disc lies in the equatorial plane of the tilted dipole component.
Warped inner edges of circumstellar discs have been shown to exist for \mbox{AA Tau} (\mbox{\citealt{1999A&A...349..619B}}; \mbox{\citealt{2003A&A...409..169B}}), and there is evidence for the existence of warps in the disc of \mbox{TW Hya} (\mbox{\citealt{2005ApJ...622.1171R}}; \mbox{\citealt{2012ApJ...757..129R}}), as well as in many other CTTS systems (\mbox{\citealt{2010A&A...519A..88A}}).
Another contributing factor to the mismatch between our models and the observationally inferred locations of accretion could be the existence of accretion streams that penetrate the stellar magnetosphere at the magnetosphere-disc boundary by the Rayleigh-Taylor instability (\mbox{\citealt{2008MNRAS.386..673K}}; \mbox{\citealt{2013MNRAS.431.2673K}}).
However, such accretion streams are likely to impact the stellar surface at low latitude, and are therefore unlikely to account for the discrepancies between observations and models.

\section{Summary and Discussion} \label{sect:summary}

\begin{figure*}
\fbox{\includegraphics[trim=0cm 1cm 0cm 0cm, clip=true, width=1.0\textwidth]{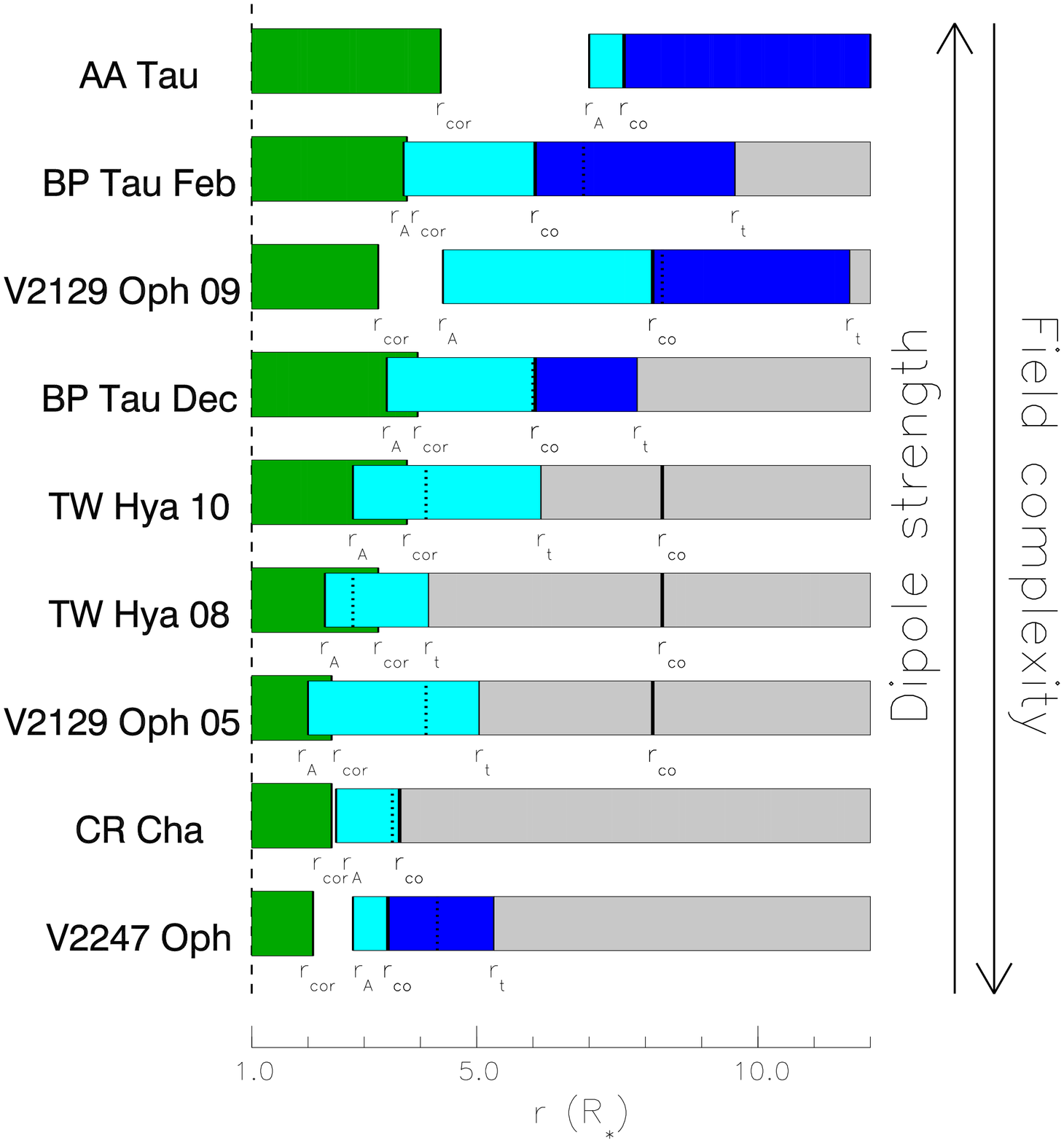}}
\caption{
Cartoon showing, for most of the stars in the sample, the coronal extents and the disc truncation radii calculated in \mbox{Sections \ref{sect:corona} and \ref{sect:acc}}.
The stars are ordered from top to bottom by decreasing strength of the dipole component of the field.
The horizontal direction is measured in stellar radii from the stellar surface, marked by the dashed line, to 12 R$_\star$.
The torque balance radius for \mbox{AA Tau} is to the right of the plot at approximately 18 R$_\star$.
The extent of the closed corona, labelled $r_{cor}$, is shown in green.
For each star, the light blue bar shows the range of radii from which accretion can occur onto the star. 
This is between the inner edge of the disc, set to the Alfv\'{e}n surface radius, and the torque balance radius, unless the torque balance radius lies outside the corotation radius, in which case it is assumed to be the corotation radius. 
For each star, the dark blue bar shows the range of radii where the magnetic field can disrupt the disc outside the corotation radius.
The vertical dotted lines show the locations of the disc truncation radii calculated using the alternative formula of \mbox{\citet{2008A&A...478..155B}}, given by \mbox{Eqn. \ref{eqn:bessolaz}}.
As the dipole components weaken, the outer edge of the corona and the inner edge of the disc moves inwards. 
}
\label{fig:radiiresults}
\end{figure*}

In this paper, we analyse the published sample of ZDI magnetic maps and use them to model the X-ray emitting closed coronae and accretion geometries for most of the stars in the sample. 
Our main results are as follows:

\begin{itemize}

\item 
Magnetic field complexity is strongly correlated with field strength.
Simple fields are the strongest, and more complex magnetic fields are the weakest. 
These properties are also related to the stellar rotation rate, with the strong simple fields being on the slowest rotators and the weak complex fields being on the most rapid rotators.
This could be a result of magnetic torques on the star from circumstellar discs, or it could be a result of stellar spin-up due to contraction and the evolution of the internal structures of PMS stars.
There is a positive correlation between rotation period and dipole field strength in the sample of CTTSs (see \mbox{Fig. \ref{fig:BdipvsP}}) with slower rotators having weaker dipole fields.
This leads to an interesting comparison with the results of the ZDI survey of young main-sequence M dwarfs where such a clear correlation is not present (\mbox{\citealt{2010MNRAS.407.2269M}}).

\item
Although most coronal X-ray emission probably originates in small-scale field structures not present in the ZDI maps, it is possible to account for the levels of X-ray emission from all of the stars in the sample using magnetically confined coronae based on the magnetic fields seen in ZDI magnetograms. 
These magnetic fields are able to hold onto the hot coronal plasma out to several stellar radii from the stellar surface, and the closed coronal extent is a strong function of the complexity of the field.
The simple dipolar fields hold onto the hot coronal plasma further from the star than the more complex fields.

\item
Typical emission measure weighted average electron densities of the coronae are between \mbox{$10^{8.8}$ cm$^{-3}$} and \mbox{$10^{10.0}$ cm$^{-3}$}.
The average electron densities in the sample are related to the field structure, with simple fields leading to lower values than complex fields.

\item 
The open unsigned magnetic flux is determined by the strength of the dipole component of the field and the stellar surface area.
Stars with stronger dipoles and larger surface areas have larger open fluxes. 
We note that we do not consider the opening up of field lines due to the action of the disc.
The locations of open and closed magnetic field is a function of field complexity.
Simple dipole fields have open field located entirely at high latitudes, covering 10-20\% of the stellar surface.
More complex fields show open field distributed over all latitudes, covering approximately 20\% of the stellar surface. 
Fields that are dominated by large-scale octupole components tend to have regions of open field at mid-latitudes extending in bands around the star, with only 5-10\% of the stellar surface covered in regions of open field.

\item
The circumstellar discs of CTTSs are typically truncated by the stellar magnetic fields several stellar radii from the stars.
In all cases, this is inside the corotation radius, though it is possible for stars with stronger dipole magnetic fields to truncate their discs outside the corotation radius, and therefore be in the propeller regime. 
Apart from the stellar mass and mass accretion rate, the location of the truncation radius is primarily determined by the strength of the dipole component of the field.
The tilt of the dipole component and the presence of higher order spherical harmonic components, can also be significant, though approximate estimates of the truncation radii can be obtained without considering these factors.

\item
The trajectories of accreting material, and the predicted locations of accretion footpoints are strongly dependent on the complexity of the star's magnetic field.
When the field is approximately dipolar, accreting material falls along large dipole-like magnetic loops and impacts the surface at high latitudes. 
For more complex fields, accretion impacts the surface at a range of latitudes. 
Several previous studies have shown that accretion footpoint locations are highly sensitive to magnetic field structure (\mbox{\citealt{2004ApJ...610..920R}}; \mbox{\citealt{2006MNRAS.371..999G}}; \mbox{\citealt{2008MNRAS.389.1839G}}; \mbox{\citealt{2008MNRAS.386.1274L}}; \mbox{\citealt{2008ApJ...687.1323M}}). 

\item
In some cases, the magnetic field is not able to hold onto the hot coronal plasma out to the radius at which the magnetic field is strong enough to disrupt the inner edge of the accretion disc.
This means that the assumption that the magnetic field structure is closed out to the inner edge of the star's accretion disc may not be justified.
It is unclear in these cases where the inner edge of the disc would lie, and what the magnetic and accretion geometries would be. 
Exploring this question further would require more complex models involving 3D magnetohydrodynamic simulations.

\end{itemize}

The coronal extents, Alfv\'{e}n surface radii, and torque balance radii are summarised in \mbox{Fig. \ref{fig:radiiresults}}.
We can see that they are strongly dependent on the strength of the dipole component of the field and the field complexity.
As the dipole component decreases in strength, and the field complexity increases, the outer edge of the closed corona, the inner edge of the disc, and the region in which the star's magnetic field is able to disrupt the disc, all move closer to the star.
If CTTSs undergo solar-like magnetic cycles, we would expect cycles to exist in the position of the inner edge of the accretion disc and the locations of accretion footpoints.

Although there is still only a small number of stars in the sample, it is interesting to try to interpret these results in terms of the evolution of stars on the PMS. 
PMS stars are initially fully convective, and as they age, they develop radiative cores if they are massive enough.
While they are still fully convective, they host strong magnetic fields with simple structures.
As they develop radiative cores, the strengths of the dipole components decay, and the fields become weaker and more complex (\mbox{\citealt{2011MNRAS.417..472D}}; \mbox{\citealt{2012ApJ...755...97G}}). 
This corresponds to a decrease in the coronal extent, with fully convective stars having large extended coronae and more evolved stars having much more compact coronae.
Observationally, this might be seen as an increase in electron densities derived from X-ray emission lines formed primarily in hot coronal plasma (a lack of such a correlation might imply that, as on the Sun, coronal X-ray emission is dominated by small-scale field structures not so far detected using the ZDI technique). 
Since the level of open unsigned magnetic flux is determined by the strength of the dipole component of the field and the stellar surface area, as PMS contract and develop radiative cores, the rate at which angular momentum is lost due to stellar winds could decrease. 
This will also correspond to a change in the structure of these winds as the latitudinal distributions of open field is a strong function of field complexity. 
For stars that develop radiative cores before their discs have dispersed, as the star evolves, and the strength of the dipole component decays, the inner edge of the accretion disc moves inwards, and the patterns of accretion footpoints become significantly more complex.

\section{Acknowledgments} 

We acknowledge the useful suggestions of the referee which lead to several improvements in the paper. 
CPJ acknowledges support from an STFC studentship. 
SGG acknowledges support from the Science \& Technology Facilities Council (STFC) via an Ernest Rutherford Fellowship [ST/J003255/1].

\appendix

\section{A Test of the Effects of Missing Small-Scale Field Structures on the Coronal Models} \label{appendix:smallscalefield}

\begin{figure} 
\includegraphics[trim= 0cm 0cm 1cm 0cm, width=0.36\textwidth]{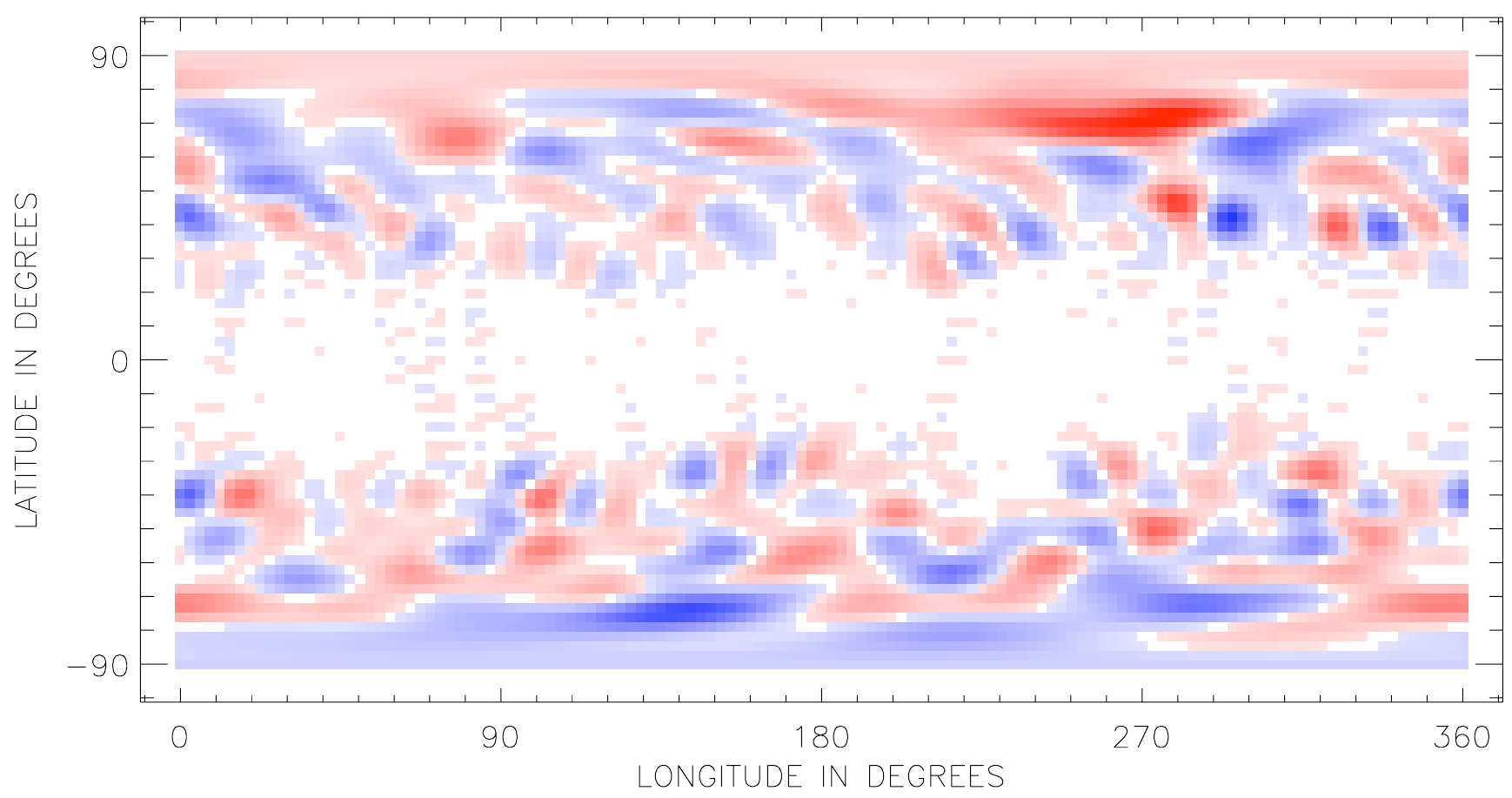}
\includegraphics[trim=2cm -0.2cm 1.2cm 1cm, clip=true, width=0.08\textwidth]{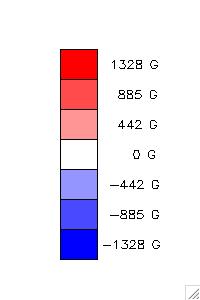}
\includegraphics[trim= 0cm 0cm 1cm 0cm, width=0.36\textwidth]{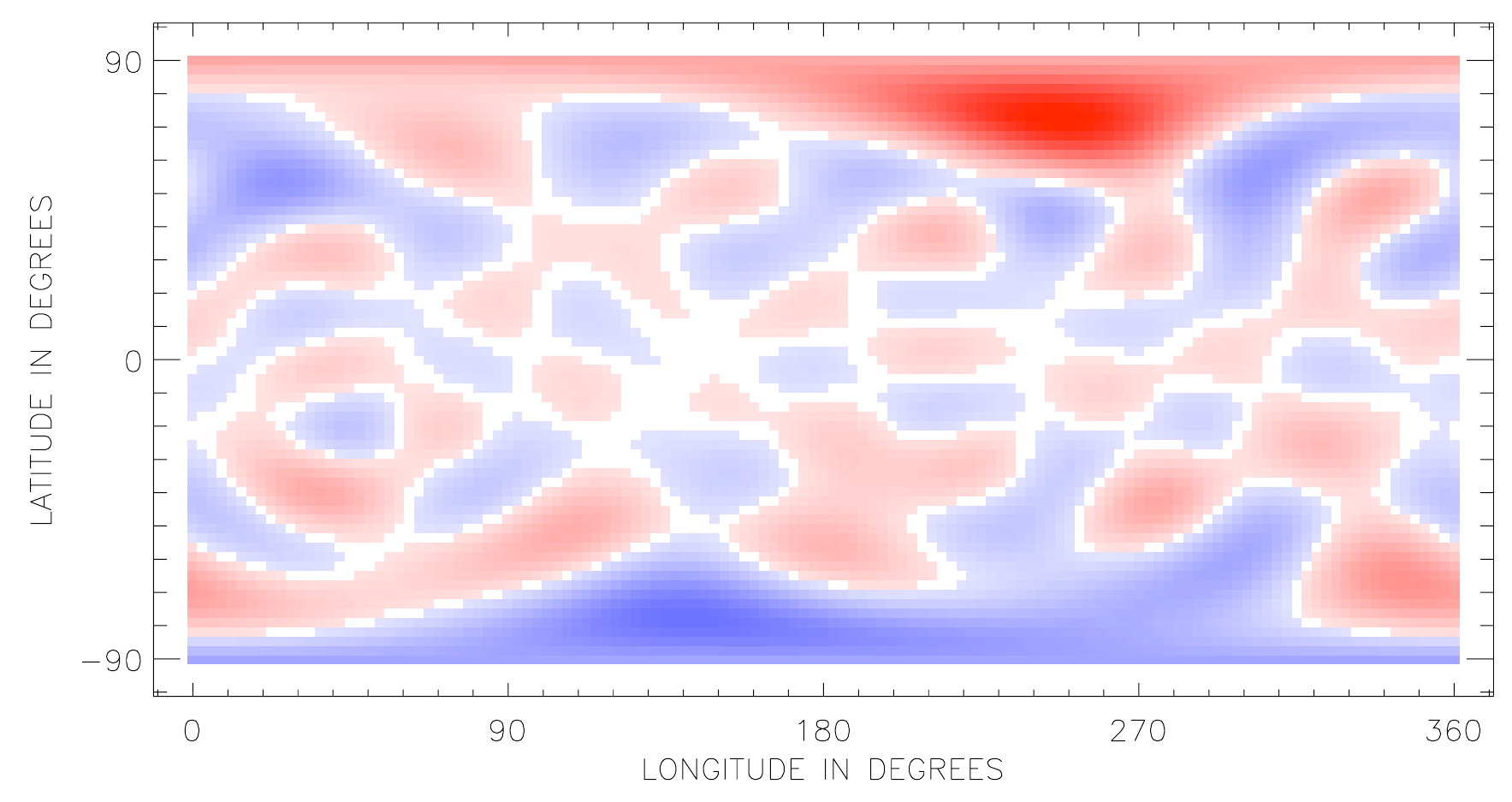}
\includegraphics[trim=2cm -0.2cm 1.2cm 1cm, clip=true, width=0.08\textwidth]{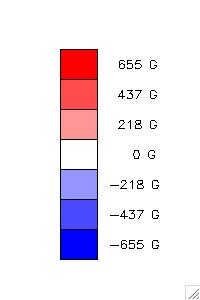}
\includegraphics[trim= 0cm 0cm 1cm 0cm, width=0.36\textwidth]{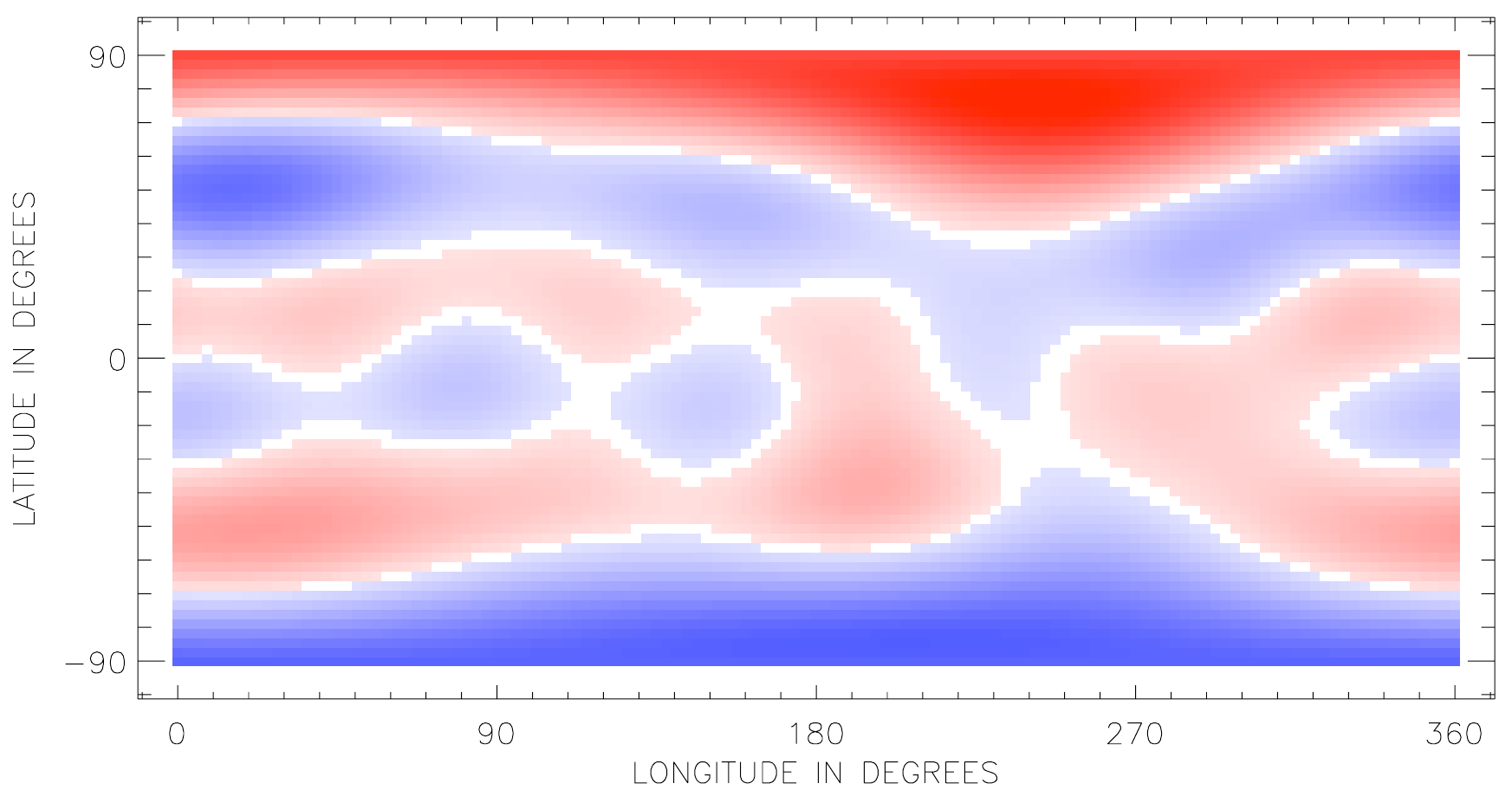} \hspace{0.5cm}
\includegraphics[trim=2cm -0.2cm 1.2cm 1cm, clip=true, width=0.08\textwidth]{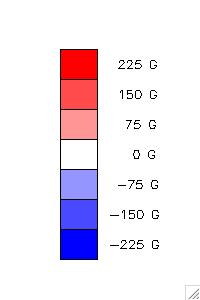}
\caption{
Radial field magnetic maps used in the tests. 
From top to bottom, the maps correspond to the original map, the map with an $l_{max}$ of 10 and the map with an $l_{max}$ of 5.
}
 \label{fig:missingfluxtestmaps}
\end{figure}

\begin{figure}
\centering
\includegraphics[trim= 0cm 0cm 1cm 0.5cm, width=0.36\textwidth]{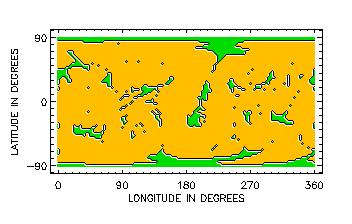}
\includegraphics[trim= 0cm 0cm 1cm 0.5cm, width=0.36\textwidth]{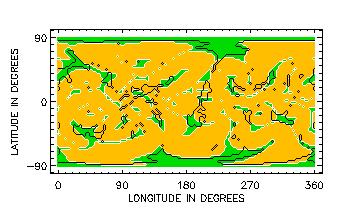}
\includegraphics[trim= 0cm 0cm 1cm 0.5cm, width=0.36\textwidth]{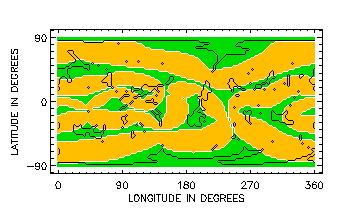}
\caption{
Similar to \mbox{Fig. \ref{fig:CTTSaccdisp}}, these maps show the locations of open (green) and closed (yellow) magnetic field structures for the three test maps.
From top to bottom, the maps correspond to the high-, medium-, and low-resolution cases respectively. 
The black contour lines show the locations of the boundaries between regions of open and closed magnetic field structures in the high-resolution case. 
The reduction in the resolution of the maps clearly leads to changes in the locations of open and closed field structures and an increase in the open flux filling factors. 
}
 \label{fig:missingfluxtestopenclosed}
\end{figure}

\begin{table}
\centering
\begin{tabular}{cccc}
 \hline
 $l_{max}$ & 30 & 10 & 5 \\
 \hline
 $\Phi_{total}$ ($10^{23}$ Mx) & 68.9 & 40.0 & 21.5 \\
 $<B>$ (G) & 120 & 70 & 40 \\
 $\log K$ & -6.2 & -4.6 & -3.9 \\
  $r_{cor}$ (R$_\star$) & 2.1 & 1.6 & 1.3 \\
 $\log n_e$ (cm$^{-3}$) & 9.2 & 9.9 & 9.9  \\
 $\Phi_{open}$  ($10^{23}$ Mx) & 9.7  & 13.0  & 11.3 \\
$f_{open}$ & 0.10 & 0.17 & 0.31 \\
$<|B_{r,open}|>$ (G) & 170 & 130 & 60 \\
 \hline
\end{tabular}
\caption{
Results of the tests of the effects of missing small-scale field structures.
The rows correspond to the maximum $l$-value spherical harmonic component used to describe the map, the total unsigned magnetic flux in the map, the surface averaged magnetic field strength, the $K$ parameter in the plasma model in \mbox{Eqn. \ref{eqn:pressurescaling}}, the coronal extent, the emission measure weighted average electron density, the total open flux, and the fraction of the stellar surface covered in open flux.
}
\label{tbl:missingfluxtestresults}
\end{table}

In \mbox{Section \ref{sect:corona}}, we present a coronal magnetic field and plasma model that uses ZDI magnetic maps to calculate average coronal electron densities, closed coronal extents, open magnetic fluxes, and the locations of open and closed magnetic field structures. 
It is well known that ZDI maps are likely to be missing magnetic flux in small-scale field structures.
In this appendix, we test the influence of this missing magnetic flux on the results of \mbox{Section \ref{sect:corona}}. 
For this, we use a simulated magnetic map produced by \mbox{\citet{2004MNRAS.354..737M}} applied the solar surface magnetic flux transport model to younger stars with enhanced meridional flow and higher latitudes of flux emergence.
The same map was used in similar tests of the influences of missing magnetic flux by \mbox{\citet{2010MNRAS.404..101J}}.
The map, shown in the upper panel of \mbox{Fig. \ref{fig:missingfluxtestmaps}}, contains many small-scale field structures that would not be reproduced in most ZDI maps.

We simulate the limited-resolution of the ZDI technique by removing all spherical harmonic components above a chosen value of $l_{max}$. 
We perform each test for $l_{max}$ values of 30 (full-resolution case), 10 (medium-resolution case) and 5 (low-resolution case), corresponding to all information reproduced in the map, a high-resolution ZDI map, and a low-resolution ZDI map respectively. 
The medium- and low-resolution cases correspond to the upper and lower limits of the resolutions of the ZDI maps used in this paper.  
For each map, we fit the values of the $K$ parameter in the plasma model (see \mbox{Eqn. \ref{eqn:pressurescaling}}) by assuming a coronal emission measure of $10^{52}$ cm$^{-3}$.
Further testing has shown that the conclusions of these tests are not sensitive to the value of the assumed emission measure. 
We assume a solar mass and radius for the star, a rotation period of 5 days, and a source-surface radius of approximately 3 R$_\star$. 

The medium- and low-resolution maps are shown in \mbox{Fig. \ref{fig:missingfluxtestmaps}}. 
Although a lot of the smallest scale magnetic field structures are missing, the medium-resolution map still contains a large amount of magnetic structure.
However, the strongest field strengths in the map, contained within the two high-latitude activity belts, are lost, leading to a decrease in the total unsigned magnetic flux. 
The low-resolution map has lost all small-scale field structures and has much lower field strengths. 

The results of the coronal models are summarised in \mbox{Table \ref{tbl:missingfluxtestresults}}.
The decreases in magnetic fluxes are accompanied by large increases in the values of the $K$ parameter. 
This is due to our assumption that at the base of the corona, the thermal pressure is given by $p_0 = K B_0^2$.
Therefore, when the magnetic flux is reduced, a larger value of $K$ is required for the coronal model to give the same coronal emission measure.
 
The large increases in the value of $K$ in the medium- and low-resolution cases correspond to larger plasma $\beta$ values at the bases of each magnetic field line. 
This means that the largest closed field lines in the high-resolution case are no longer able to hold onto the coronal plasma and are blown open, which leads to a reduction in our calculated coronal extents.
The emission measure weighted average election densities are increased by about a factor of five when we decrease the resolutions of the maps, though we find no significant difference in the calculated values between the medium- and low-resolution cases. 

In \mbox{Table \ref{tbl:missingfluxtestresults}} and \mbox{Fig. \ref{fig:missingfluxtestopenclosed}}, we show the effects on the calculated open fluxes and the predicted locations of open and closed magnetic field structures.
As the resolution of the map is reduced, the strengths of the magnetic fields in regions of open field are also reduced.
The magnitude of this reduction is similar to the magnitude of the reduction in the total unsigned magnetic flux.
Reducing the resolution of the map leads to a large increase in the open flux filling factor, which is clearly shown in \mbox{Fig. \ref{fig:missingfluxtestopenclosed}}, and changes to the locations of open field.
There is a good match between the locations of open field in the full-resolution case and the medium-resolution case.
For the low-resolution case, most of the largest open field regions in the high-resolution case are reproduced, but there are new regions of open field that are not present in the high-resolution case. 
The increases in the open flux filling factors are similar to the decreases in the magnetic field strengths in regions of open field, and therefore, the calculated open fluxes are not affected when the resolutions of the maps are reduced.
This is consistent with the conclusion that a star's open flux is determined primarily by the strength of the dipole component of the field (see \mbox{Fig. \ref{fig:openfluxresults}}) and the stellar radius. 

We have shown that the exclusion of small-scale magnetic field structures in magnetic maps has a modest effect on the coronal plasma model. 
It may lead to a small decrease in both the coronal size and an increase in the inferred coronal densities, but both of these effects are well within the current observational uncertainties.

\bibliographystyle{mn2e}
\bibliography{mybib}

\end{document}